\documentclass[conference]{IEEEtran}
\IEEEoverridecommandlockouts
\usepackage{cite}
\usepackage{amsmath,amssymb,amsfonts}
\usepackage{algorithmic}
\usepackage{graphicx}
\usepackage{textcomp}
\usepackage{xcolor}
\usepackage{hyperref}
\usepackage{booktabs,tabularx}
\usepackage{braket}

\usepackage{orcidlink}

\newcommand{\od}{\odot}

\renewcommand{\figureautorefname}{Figure~\negthinspace}
\renewcommand{\equationautorefname}{Equation~\negthinspace}

\def\BibTeX{{\rm B\kern-.05em{\sc i\kern-.025em b}\kern-.08em
    T\kern-.1667em\lower.7ex\hbox{E}\kern-.125emX}}
\begin{document}

\title{Self-Modulating Quantum Fast-Weight Programmers for Efficient Adaptive Sequential Learning
\thanks{
The views expressed in this article are those of the authors and do not represent the views of Wells Fargo. This article is for informational purposes only. Nothing contained in this article should be construed as investment advice. Wells Fargo makes no express or implied warranties and expressly disclaims all legal, tax, and accounting implications related to this article.}
}

\author{Samuel Yen-Chi Chen$^1$\orcidlink{0000-0003-0114-4826}, Yifeng Peng$^2$\orcidlink{0009-0007-3306-9417}, Kuo-Chung Peng$^6$\orcidlink{0009-0001-8342-2481}, Jiun-Cheng Jiang$^6$\orcidlink{0009-0005-1134-4962}, Chun-Hua Lin$^6$\orcidlink{0009-0002-4383-0453}\\Junghoon Justin Park$^3$\orcidlink{0000-0001-8982-0387}, Huan-Hsin Tseng$^4$\orcidlink{0000-0001-9544-4226}, Hsin-Yi Lin$^4$\orcidlink{0000-0001-5731-2353}
Kuan-Cheng Chen$^5$\orcidlink{0000-0002-6575-7034}, Chen-Yu Liu$^6$\orcidlink{0000-0002-5437-5188}, Shinjae Yoo$^4$\orcidlink{0000-0003-4378-6448}\\
\small $^1$Wells Fargo
\small $^2$Stevens Institute of Technology
\small $^3$Seoul National University \\
\small $^4$Brookhaven National Laboratory
\small $^5$Imperial College London 
\small $^6$National Taiwan University\\
}

\maketitle

\begin{abstract}
Recent advances in quantum machine learning have motivated efficient models for sequential data processing. In this paper, we propose Self-Modulating Quantum Fast Weight Programmers, or Self-Modulating QFWP, which extends Quantum Fast Weight Programmers by introducing adaptive modulation over both newly generated fast-weight updates and historical fast-weight memory. Numerical results show that the proposed mechanism improves convergence stability and prediction performance across varying model settings, including different numbers of qubits and input sequence lengths. We further provide theoretical arguments explaining how self-modulation balances new information injection with memory retention, thereby enhancing temporal information propagation. These results suggest that Self-Modulating QFWP is a compact and effective framework for quantum machine learning on time-series data.
\end{abstract}

\begin{IEEEkeywords}
Quantum Neural Networks, Variational Quantum Circuits, Meta-Learning, Learning to Learn, Sequence Learning
\end{IEEEkeywords}

\section{Introduction}
Quantum computing (QC) provides new computational paradigms that may offer advantages for certain problems \cite{nielsen2010quantum,shor1994algorithms,grover1996fast}. Combined with the success of artificial intelligence and machine learning (AI/ML), this has motivated the development of quantum machine learning (QML), which seeks to design learning models based on quantum computational principles \cite{biamonte2017quantum,chen2026qai,delgado2025quantum,schuld2019quantum,benedetti2019parameterized,abbas2021power,sim2019expressibility}. A central framework in QML is the variational quantum algorithm (VQA) \cite{bharti2022noisy,cerezo2021variational,peruzzo2014variational,wang2021noise}, which has been widely used in hybrid quantum-classical models for classification, sequence learning, and reinforcement learning \cite{mitarai2018quantum,schuld2020circuit,chen2022quantumLSTM,chen2020QRL,hur2022quantum,macaluso2022variational,marchisio2025cutting,mujal2023time,kutvonen2020optimizing,chen2025validating,chen2025quantum,liu2024qtrl,liu2025quantum,chen2025qtransformer,cong2019quantum,skolik2022quantum,jerbi2021parametrized,kundu2024enhancing,kundu2024reinforcement,sequeira2023policy}.

Sequence learning is an important direction in QML because many real-world data are naturally temporal. Existing quantum recurrent models, such as Quantum Long Short-Term Memory (QLSTM) networks, have shown that parameterized quantum circuits can be used to process sequential information \cite{chen2022quantumLSTM,li2023pqlm,di2022dawn,stein2023applying,takaki2021learning,bausch2020recurrent,siemaszko2023rapid,li2023quantum,viqueira2025density,hsu2025quantum,liu2025federated,lin2024quantum,chen2025rwkv}. However, many recurrent quantum models rely on fixed temporal update structures, leaving open how quantum architectures should regulate the interaction between newly generated information and memory accumulated from previous time steps.

Quantum Fast Weight Programmers (QFWPs) provide an alternative approach to quantum sequence modeling by dynamically generating fast-weight parameters during recurrent processing \cite{chen2024QFWP}. This mechanism allows the model to encode temporal information into evolving fast-weight states. Nevertheless, directly accumulating fast-weight updates may cause unstable parameter evolution or inefficient memory propagation, especially when the input sequence length or model size changes. Therefore, an explicit mechanism for controlling both new fast-weight updates and historical fast-weight memory is needed.

To address this issue, we propose Self-Modulating Quantum Fast Weight Programmers, or Self-Modulating QFWP. The key idea is to introduce adaptive modulation into the fast-weight evolution, allowing the model to regulate how much newly generated information is injected and how much historical fast-weight memory is retained. This self-modulating mechanism provides a more flexible recurrent structure and aims to improve learning robustness across different quantum model settings.

\textbf{Contribution Statement.} The main contributions of this work are threefold. First, we propose a self-modulating extension of QFWP that adaptively regulates fast-weight dynamics for more robust sequence learning. Second, we conduct comprehensive numerical simulations and ablation studies on multiple time-series tasks, evaluating the effects of different numbers of qubits, input sequence lengths, and modulation designs. Third, we provide theoretical arguments explaining why certain forms of self-modulation more effectively balance new information injection and memory retention, thereby improving temporal information propagation.

\section{Related Works}
\label{sec:related_works}

Sequence learning is a central topic in both classical and quantum machine learning. 
In the classical setting, the long short-term memory (LSTM) network~\cite{hochreiter1997long} has long served as the canonical recurrent architecture, using input, forget, and output gates to mitigate vanishing gradients and capture long-range dependencies. 
Building on this design, the Quantum LSTM (QLSTM)~\cite{chen2022quantumLSTM} replaces the classical components inside the LSTM cell with variational quantum circuits (VQCs), and has since been applied across natural language processing~\cite{li2023pqlm,di2022dawn,stein2023applying}, generative modeling~\cite{chu2025lstm}, reinforcement learning~\cite{chen2023quantumLSTM_RL,chen2024quantumLSTM_RC_RL}, time-series forecasting~\cite{cao2023_linear_enahnced_QLSTM,chen2025benchmarking_SMS_app}, solving partial differential equations~\cite{chen2025_QLSTM_for_PDE}, and many other domains~\cite{tripathi2025_QLSTM_DDoS_detection,elsayed2025hybrid,tran2025_QLSTM_IoT_Sensing,chen2025_QLSTM_CNC,11196477_QLSTM_WIND_TURBINE_FAULT_DETECTION,chen2025_MultiChip_QLSTM,hsu2025qkanlstm,hsu2025_federated_QKLSTM_HAR,rosato2025study_QLSTM_RC_AUTOENCODER}. 
Recent QKAN-LSTM work further explores quantum-inspired Kolmogorov--Arnold recurrent gates to achieve better scalability~\cite{hsu2025qkanlstm,jiang2025qkan}.
However, QLSTM inherits the structural drawbacks of recurrent designs: its non-linear gated recurrence couples the hidden state across all time steps, forcing gradient computation to proceed sequentially through backpropagation-through-time (BPTT) and precluding parallelization across time steps. 
This sequential cost is amplified in the quantum setting, where each step requires repeated VQC evaluations for gradient estimation.

The Fast Weight Programmer (FWP) framework~\cite{schmidhuber1992learning,schmidhuber1993reducing} addresses sequence learning through a different mechanism: a slow network generates low-rank parameter updates that reprogram a fast network, yielding a linear recurrence in the fast weights. This formulation was later shown to be equivalent to linearized self-attention~\cite{schlag2021linear,katharopoulos2020transformers}, and extended through recurrent and self-referential variants~\cite{irie2021going,irie2023practical}. The Quantum FWP (QFWP)~\cite{chen2024QFWP} adapts this paradigm to the hybrid setting, with a classical slow programmer continually reprogramming a VQC fast programmer. Because temporal dependence is encoded through additive parameter updates rather than a recurrent quantum state, the QFWP avoids deep gradient evaluation such as BPTT through the quantum circuit. Subsequent variants such as the Quantum-Train QFWP~\cite{liu2025programming} and observable-aware extensions~\cite{chen2025learning_observables} further reduce parameters and broaden applicability. However, these models share a purely additive update rule that lacks any input-dependent mechanism to forget or reweight accumulated information, in contrast to QLSTM. To close this gap, the Self-Modulating QFWP introduces input-dependent modulation of both the new update and the previously accumulated fast weights. Because the modulation coefficients depend only on the current input, the recurrence remains linear and parallelizable~\cite{martin2018parallelizing}, thereby inheriting the temporal-memory control of LSTM-style temporal-memory control without the sequential gradient bottleneck of non-linear recurrence.

\section{Quantum Neural Networks}
A quantum neural network (QNN), also known as variational quantum circuit (VQC) or parameterized quantum circuit (PQC), consists of a data-encoding circuit $U(\vec{x})$, a trainable variational circuit $W(\Theta)$, and a final measurement stage. Given a classical input $\vec{x}$, the quantum state is prepared as
$
\ket{\Psi}=W(\Theta)U(\vec{x})\ket{0}^{\otimes n},
$
where $n$ is the number of qubits and $\Theta$ denotes the trainable circuit parameters. The model output is obtained by measuring Hermitian observables $\hat{B}_k$, yielding expectation values
$
\langle \hat{B}_k\rangle=\bra{\Psi}\hat{B}_k\ket{\Psi}.
$
Therefore, the QNN defines a quantum function $\vec{f}(\vec{x};\Theta)=(\langle \hat{B}_1\rangle,\dots,\langle \hat{B}_K\rangle)$.

\section{Quantum Fast Weight Programmers and Learning to Modulate}

Let $L$ be the number of trainable variational layers in the fast quantum circuit and $Q$ be the number of qubits. At time step $t$, the fast circuit state is $\Theta_t\in\mathbb{R}^{L\times Q}$, where $(\Theta_t)_{k,q}$ is the circuit angle assigned to layer $k\in\{1,\ldots,L\}$ and qubit $q\in\{1,\ldots,Q\}$. 

Denote $\Omega$ as the collection of trainable parameters of the classical slow controller. 
These parameters are optimized during training and remain fixed during inference. 
By contrast, $\Theta_t$, $\Delta_t$, $M_t^{\mathrm{new}}$, and $M_t^{\mathrm{old}}$ are time-dependent quantities generated from the input sequence. 

Let $h_t = \phi_{\Omega_\phi}(x_t)\in\mathbb{R}^{H}$ be the hidden state produced by the classical controller $\phi_{\Omega_\phi}$ from the input $x_t$. The raw update vectors are $\ell_t = W_{\ell} \, h_t + b_{\ell} \in \mathbb{R}^{L}$, $r_t = W_r \, h_t + b_r \in \mathbb{R}^Q$ and the raw fast-weight update is the rank-1 matrix $\Delta_t = \ell_t \cdot r_t^\top\in\mathbb{R}^{L\times Q}$. Then the original QFWP \cite{chen2024QFWP} uses the update rule,
\begin{equation}
    \Theta_t = \Theta_{t-1} + \Delta_t
\end{equation}

Thus, the fast circuit state is obtained by accumulating all previous raw updates. This makes the original QFWP a quantum recurrent fast-weight model: the classical slow programmer generates a low-rank update from the current input, and the quantum circuit uses the accumulated fast parameters to produce quantum expectation-value features.

\begin{figure}[htbp]
\vskip -0.1in
\centering
\includegraphics[width=1\columnwidth]{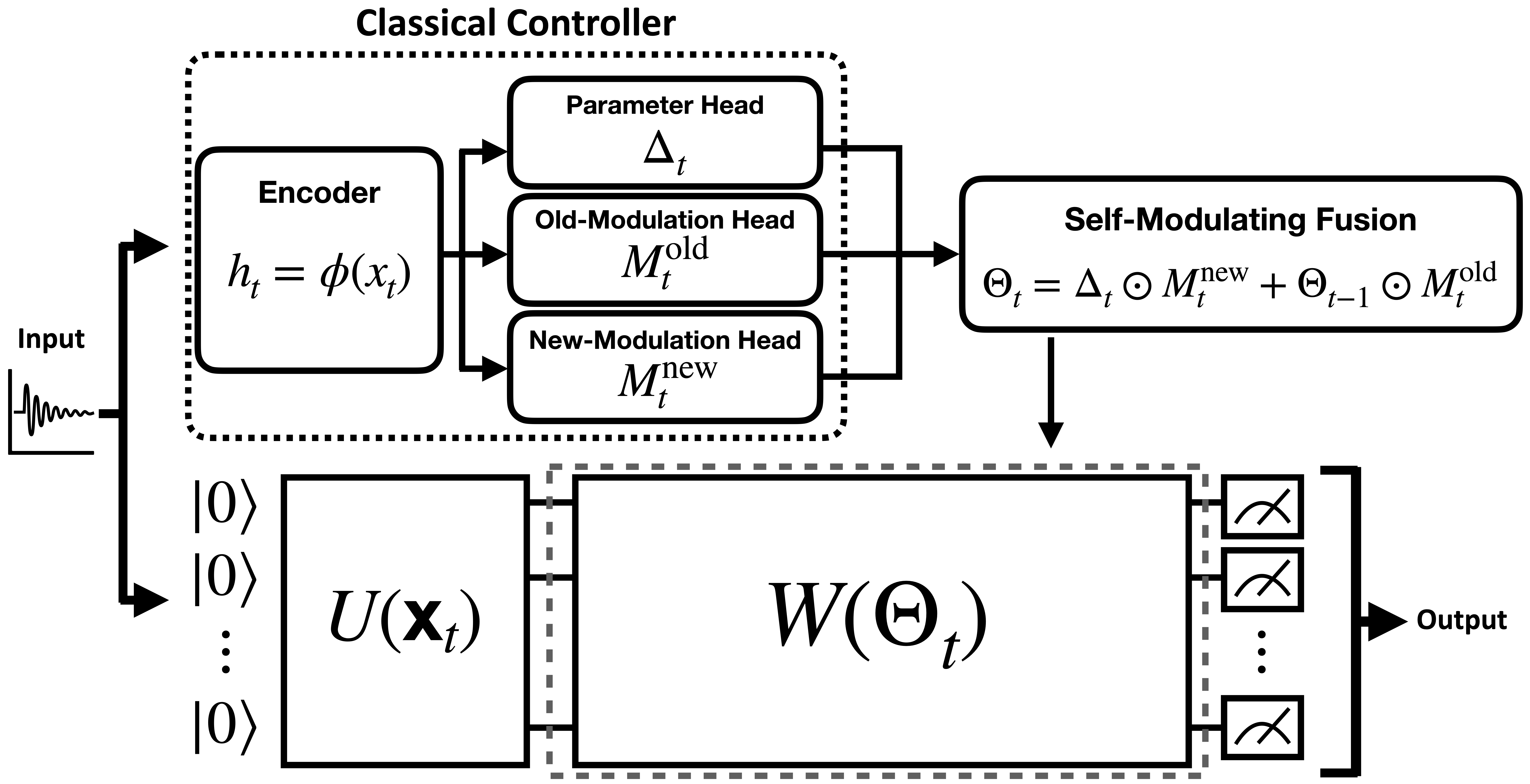}
\caption{\textbf{Architecture of the proposed Self-Modulating QFWP.} A classical controller generates $\Delta_t$, $M_t^{\mathrm{old}}$, and $M_t^{\mathrm{new}}$, which are fused to produce the circuit parameters $\Theta_t$ for the variational quantum circuit $W(\Theta_t)$.}
\label{fig:self_modulating_qfwp_diagram}
\vskip -0.15in
\end{figure}

\subsection{Full Self-Modulating QFWP}

The full Self-Modulating QFWP, as illustrated in \figureautorefname{\ref{fig:self_modulating_qfwp_diagram}}, introduces two input-dependent modulation matrices, one for the current raw update $\Delta_t$, and one for the previous fast circuit state $\Theta_{t-1}$. The new-update modulation head generates,
\[
m_t^{\{\mathrm{old, new}\},\{L,Q\} } = W_{\{\mathrm{old, new}\},\{L,Q\}} \, h_t + b_{\{\mathrm{old, new} \},\{L,Q\}} %
\]
compactly representing the following 4 vectors,
\[
m_t^{\mathrm{new},L}\in\mathbb{R}^{L}, \,\, m_t^{\mathrm{new},Q}\in\mathbb{R}^{Q}, \,\, m_t^{\mathrm{old},L}\in\mathbb{R}^{L}, \,\, m_t^{\mathrm{old},Q}\in\mathbb{R}^{Q}
\]
with the corresponding $\{\mathrm{old, new}\}$-update modulation,
\begin{equation}
    M_t^{\{\mathrm{old, new}\}} = m_t^{\{\mathrm{old, new}\},L} \left(m_t^{\{\mathrm{old, new}\},Q}\right)^\top \in\mathbb{R}^{L\times Q}
\end{equation}
We define the full self-modulating update rule as,
\begin{equation}\label{E: full modulation}
    \Theta_t = \Delta_t\odot M_t^{\mathrm{new}} + \Theta_{t-1}\odot M_t^{\mathrm{old}},
\end{equation}
where $\odot$ denotes the element-wise multiplication.

This model controls both the newly generated update and the previously accumulated fast parameters. New modulation regulates how the current update is written, while Old modulation regulates how the previous fast-weight memory is retained, suppressed, amplified, or sign-reversed. Since the modulation matrices are produced by linear layers and outer products, they are not bounded gates unless an additional bounding nonlinearity is introduced.

\subsection{Only-New Self-Modulating QFWP}

The \textbf{Only-New} variant is a special case of Eq.~(\ref{E: full modulation}) where only the new raw update is modulated, leaving the previous fast parameters unchanged by $\Theta_t=\Delta_t\od M_t^{\mathrm{new}}+\Theta_{t-1}$. This model tests whether input-dependent write modulation alone is sufficient. %

\subsection{Only-Old Self-Modulating QFWP}

On the other hand, the \textbf{Only-Old} variant $\Theta_t=\Delta_t+\Theta_{t-1}\od M_t^{\mathrm{old}}$ is considered to modulate only the previous fast parameters. This model tests whether input-dependent memory modulation alone captures most of the benefit of the full model. %

\begin{table}[t]
\centering
\caption{Comparison of update rules in QFWP variants.}
\label{tab:qfwp_rules}
\begin{tabularx}{\columnwidth}{@{}l>{\raggedright\arraybackslash}X@{}}
\toprule
\textbf{Model} & \textbf{Update Rule} \\
\midrule
Original QFWP & $\Theta_t=\Theta_{t-1}+\Delta_t$ \\
Full Self-Modulating QFWP & $\Theta_t=\Delta_t\odot M_t^{\mathrm{new}}+\Theta_{t-1}\odot M_t^{\mathrm{old}}$ \\
Only-New Self-Modulating QFWP & $\Theta_t=\Delta_t\odot M_t^{\mathrm{new}}+\Theta_{t-1}$ \\
Only-Old Self-Modulating QFWP & $\Theta_t=\Delta_t+\Theta_{t-1}\odot M_t^{\mathrm{old}}$ \\
\bottomrule
\end{tabularx}
\end{table}

\section{Numerical Results and Discussions}
We evaluate the performance of the proposed Self-Modulating-QFWP model under the following four configurations:
(1) \emph{Self-Modulating-QFWP} — this is the variant in which both new and old parameters are modulated;
(2) \emph{Self-Modulating-QFWP (Only Old)} — this is the variant in which only old parameters are modulated;
(3) \emph{Self-Modulating-QFWP (Only New)} — this is the variant in which only new parameters are modulated;
(4) \emph{QFWP (Standard)} — this is the original QFWP in which no parameter modulation is implemented.

To compare the performance difference across various model sizes and qubit counts, we consider and train the following settings for hidden size, which is also the number of qubits of VQCs in the fast programmer: 4, 6 ,8, 10, 12, 14.
The training and evaluation protocol follows the methodology described in~\cite{chen2022quantumLSTM,chen2022reservoir,chen2024QFWP,chen2025_DiffQAS_QLSTM}. Specifically, the model is trained to predict the $(N+1)$-th value in a sequence given the preceding $N$ observations. For example, at time step $t$, the input to the model is $[x_{t-4}, x_{t-3}, x_{t-2}, x_{t-1}]$ (with $N=4$), and the target output is $y_t$, which should approximate the ground truth $x_t$. To investigate the memory performance of the proposed model across various sequence length, we consider $N = \{4, 8, 16, 32, 64\}$. For each time-series task, each model configuration is evaluated over $6\times 5=30$ combinations, obtained from six hidden-size/qubit-count settings $H\in\{4,6,8,10,12,14\}$ and five input sequence lengths $N\in\{4,8,16,32,64\}$.

We consider five time-series tasks which are used as benchmarks in previous studies \cite{chen2022quantumLSTM,chen2022reservoir,chen2024QFWP,chen2025_DiffQAS_QLSTM}: \emph{Damped SHM}, \emph{Bessel function $J_2$}, \emph{Delayed Quantum Control}, \emph{NARMA-5}, and \emph{NARMA-10}. All simulations use \texttt{batch\_size}=4, number of QNN layers $L=5$, \texttt{learning\_rate}=$10^{-3}$ and the Adam optimizer.

In addition to the standard metric such as mean square error (MSE), we consider the following metrics to better capture the learning performance of different variants in our proposed Self-Modulating QFWP models. 
\begin{itemize}
    \item \textbf{Relative Improvement}: 
    \begin{equation}
        \Delta^{\mathrm{rel}} = \frac{M_{\mathrm{standard}} - M_{\mathrm{variant}}}{M_{\mathrm{standard}} + \varepsilon}.
        \label{eq:relative_improvement}
    \end{equation}
    This quantity measures how much the variant improves or degrades the metric $M$ relative to the standard baseline, while using $\varepsilon$ to prevent numerical instability when the denominator is small. When $M_{\mathrm{variant}} < M_{\mathrm{standard}}$, the numerator is positive, so $\Delta^{\mathrm{rel}} > 0$. This means that the variant achieves a smaller value of the metric and therefore performs better. Accordingly,
        \begin{itemize}
            \item \(\Delta^{\mathrm{rel}} > 0\): variant better
            \item \(\Delta^{\mathrm{rel}} < 0\): variant worse
            \item \(\Delta^{\mathrm{rel}} \approx 0\): both perform similarly
        \end{itemize}
    \item \textbf{Relative Strength}:
    We define \emph{Relative Strength} to quantify whether old-parameter modulation or new-parameter modulation contributes more strongly to performance improvement. Specifically, Relative Strength is defined as the difference between the relative improvements of the Only-Old-Modulated and Only-New-Modulated variants:
        \begin{equation}
            \mathrm{Relative\ Strength}
            =
            \Delta_{\mathrm{old}}^{\mathrm{rel}}
            -
            \Delta_{\mathrm{new}}^{\mathrm{rel}}.
            \label{eq:relative_strength}
        \end{equation}
    A positive Relative Strength value indicates that old-parameter modulation contributes more strongly than new-parameter modulation, while a negative value indicates the opposite. Values close to zero suggest that the two modulation pathways have comparable effects.
    \item \textbf{Synergy}: We further define \emph{Synergy} to measure whether jointly modulating both old and new parameters provides an additional benefit beyond the better single-sided modulation strategy. Formally, Synergy is defined as
    \begin{equation}
        \mathrm{Synergy}
        =
        \Delta_{\mathrm{both}}^{\mathrm{rel}}
        -
        \max
        \left(
        \Delta_{\mathrm{old}}^{\mathrm{rel}},
        \Delta_{\mathrm{new}}^{\mathrm{rel}}
        \right).
        \label{eq:synergy}
    \end{equation}
    A positive Synergy value indicates that jointly modulating both old and new parameters outperforms the better of the two single-sided modulation variants, suggesting a genuine cooperative effect. A value near zero indicates little additional gain beyond the stronger single-sided modulation, while a negative value suggests that the joint modulation strategy does not surpass the better individual component.
    \end{itemize}
\begin{figure}[htbp]
\vskip -0.1in
\centering
\includegraphics[width=1\columnwidth]{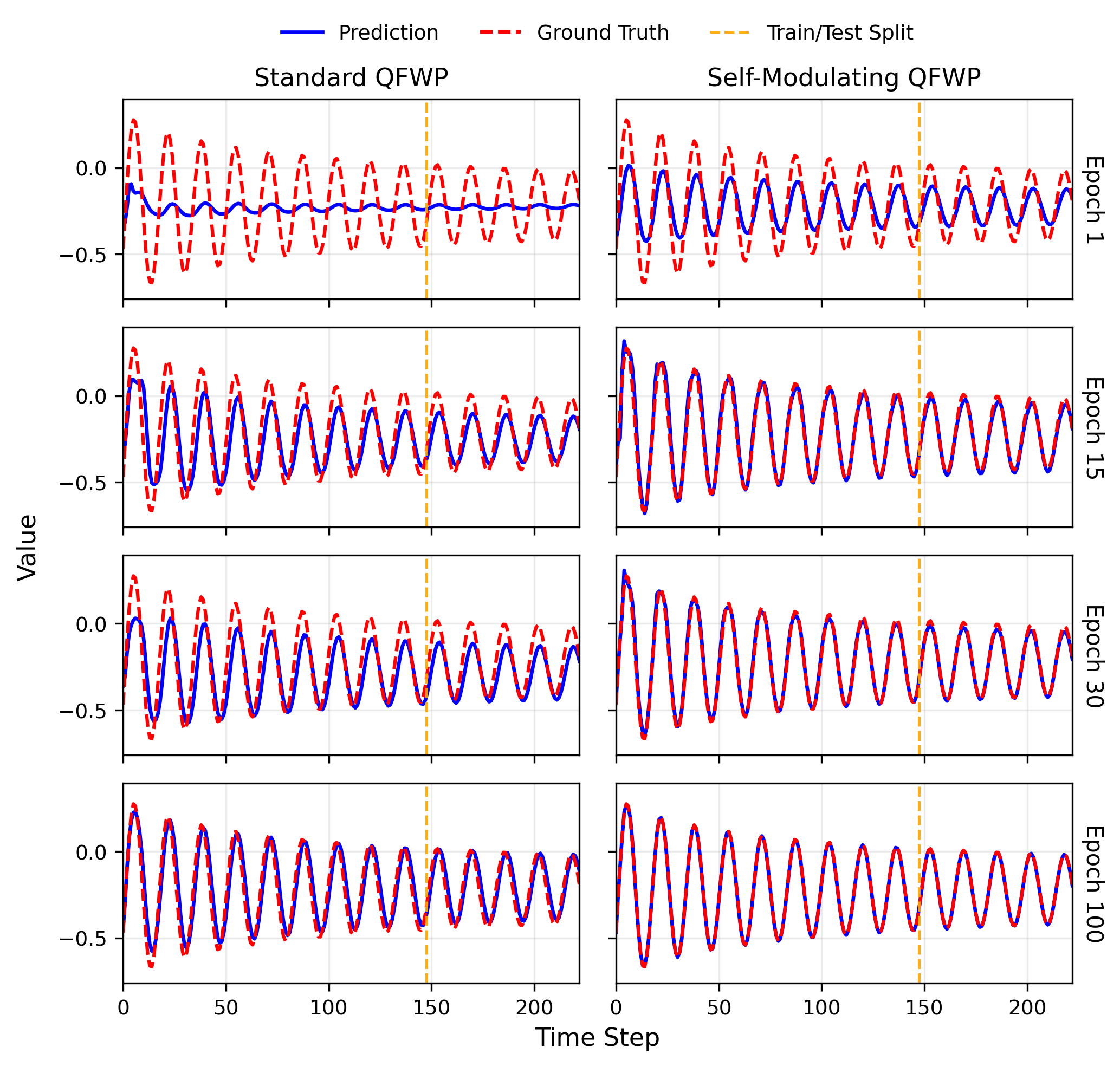}
\caption{\textbf{Prediction trajectories of Standard QFWP and full Self-Modulating QFWP on the \texttt{bessel\_j2} task at selected training epochs (\texttt{seq\_len}=32).} Blue solid lines denote model predictions, red dashed lines denote ground truth, and orange dashed lines indicate the train/test split.}
\label{fig:bessel_j2_rollout}
\vskip -0.15in
\end{figure}
For the \texttt{bessel\_j2} task with sequence length 32, as shown in \figureautorefname{\ref{fig:bessel_j2_rollout}}, the Self-Modulating QFWP exhibits a clearly faster convergence behavior than the standard QFWP. At epoch 1, the standard model still produces an almost flattened response and fails to capture the oscillatory structure of the target sequence, whereas the Self-Modulating variant already aligns reasonably well with the underlying periodicity, phase, and damping trend. By epoch 15, the prediction of the Self-Modulating QFWP nearly overlaps with the ground truth, while the standard QFWP still shows visible amplitude and local shape mismatches. This advantage remains evident in both the training and testing regions, indicating that self-modulation improves not only fitting speed but also early-stage generalization. Although both models eventually achieve strong prediction quality by epoch 100, the main benefit of the Self-Modulating QFWP in this task lies in its substantially faster and more stable learning of the correct temporal dynamics.
\begin{figure}[htbp]
\vskip -0.1in
\centering
\includegraphics[width=1\columnwidth]{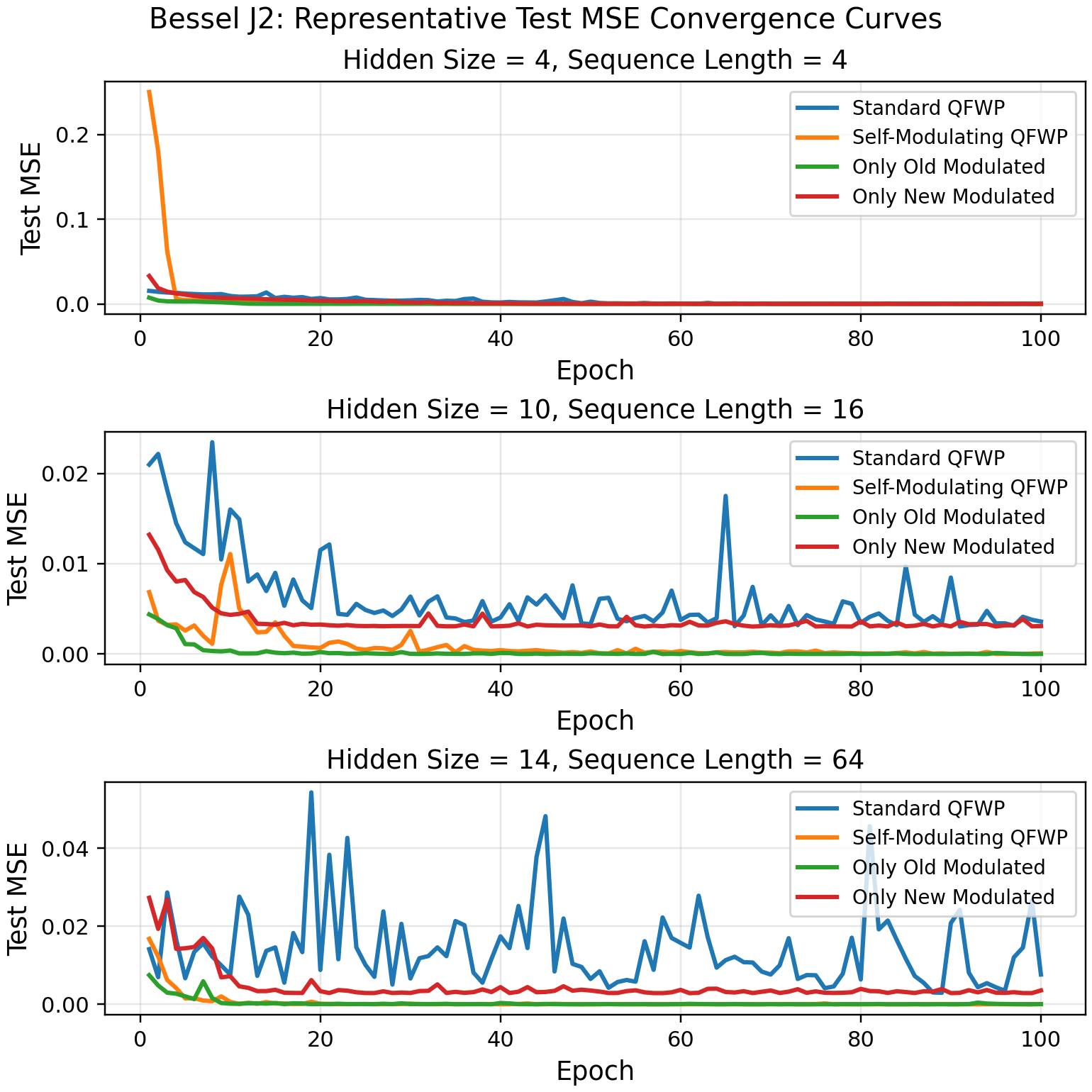}
\caption{\textbf{Representative test MSE convergence curves on the \texttt{bessel\_j2} task, comparing Standard QFWP, full Self-Modulating QFWP, and its ablation variants under selected hidden sizes and sequence lengths.}}
\label{fig:bessel_j2_convergence}
\vskip -0.15in
\end{figure}
\figureautorefname{\ref{fig:bessel_j2_convergence}} shows representative test MSE convergence curves for the \texttt{bessel\_j2} task under several hidden-size (qubit-count) and sequence-length settings. In the easiest setting (hidden size = 4, sequence length = 4), all four variants eventually converge to nearly zero test error, although the modulated variants exhibit faster and smoother early-stage optimization. As the task becomes more challenging, especially at sequence lengths 16 and 64, the differences between methods become substantially more pronounced. The standard QFWP consistently maintains higher test MSE and exhibits noticeable fluctuations throughout training. The Only New Modulated variant improves over the standard model, but typically settles at a visibly higher error floor. In contrast, both the Self-Modulating QFWP and the Only Old Modulated variant rapidly drive the test MSE to near zero and maintain much more stable convergence trajectories, with Only Old Modulated appearing particularly stable across the harder settings. These results suggest that, for oscillatory sequence modeling in \texttt{bessel\_j2}, modulation based on historical or old-state information is likely the dominant contributor to the performance gain, while full self-modulation provides similarly strong and robust optimization behavior.
\begin{figure}[htbp]
\vskip -0.1in
\centering
\includegraphics[width=1\columnwidth]{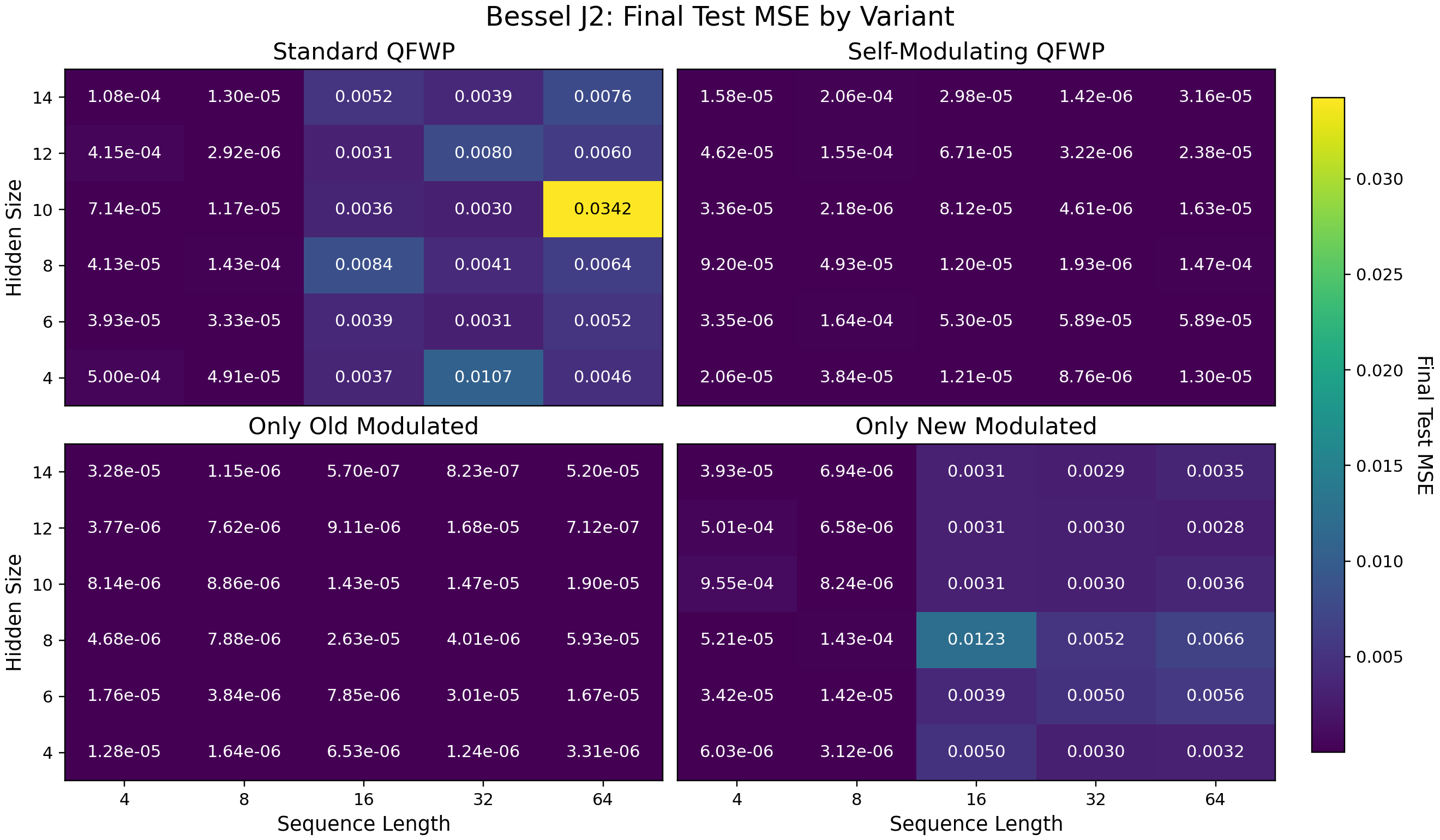}
\caption{\textbf{Final test MSE of Standard QFWP, Self-Modulating QFWP, and its ablation variants on the \texttt{bessel\_j2} task across hidden sizes and sequence lengths.} Each cell reports the final test MSE for the corresponding configuration.}
\label{fig:bessel_j2_variants_loss_comparison}
\vskip -0.12in
\end{figure}
The final-test-MSE heatmaps (\figureautorefname{\ref{fig:bessel_j2_variants_loss_comparison}}) further consolidate the trends suggested by the epoch-wise prediction plots and representative convergence curves. For the \texttt{bessel\_j2} task, the standard QFWP maintains low error only in shorter-sequence settings (e.g., sequence lengths 4 and 8), but its performance degrades substantially as the sequence length increases to 16, 32, and 64. In contrast, the Self-Modulating QFWP remains consistently strong across the entire grid of hidden sizes and sequence lengths, with final test errors staying in a uniformly low range. Even more notably, the Only Old Modulated variant appears to be the strongest overall configuration on this task, with most entries remaining at the $10^{-6}$ to $10^{-5}$ level and showing even greater consistency than the full Self-Modulating QFWP. By comparison, the Only New Modulated variant performs reasonably well for short sequences, but deteriorates markedly for longer ones, exhibiting a failure pattern much closer to the standard QFWP. Taken together, these results strongly suggest that, for the oscillatory and history-dependent dynamics of \texttt{bessel\_j2}, old-state modulation is the dominant source of performance improvement, while full self-modulation preserves this advantage and provides robust performance across configurations.
\begin{figure}[htbp]
\vskip -0.1in
\centering
\includegraphics[width=1\columnwidth]{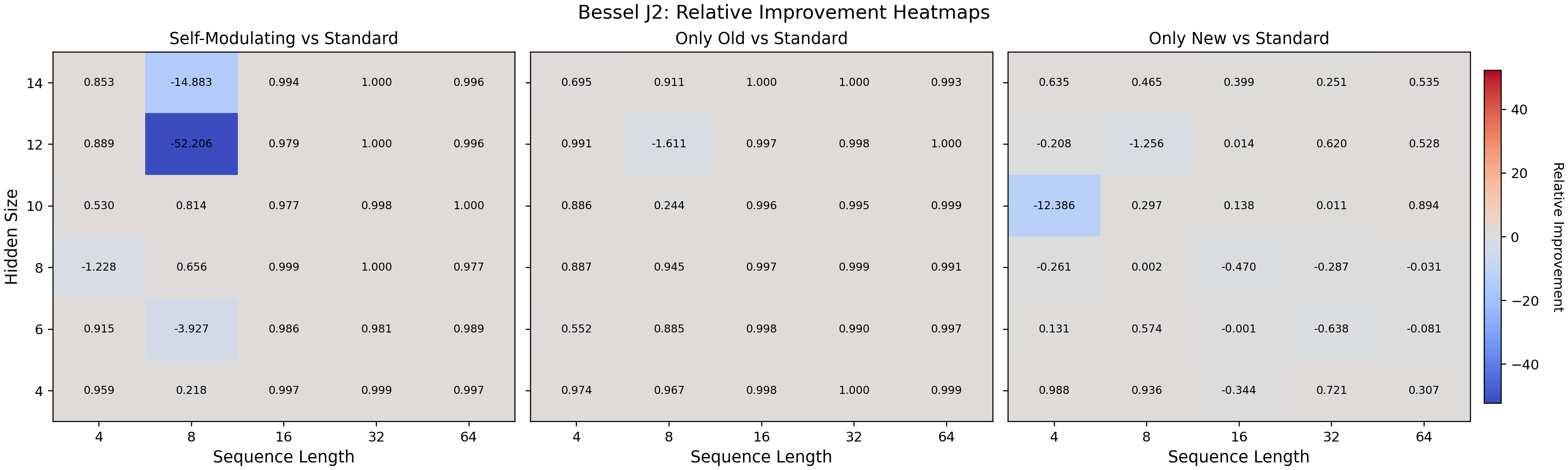}
\caption{
\textbf{Relative improvement over Standard QFWP on \texttt{bessel\_j2}.} Positive values indicate improvement; a few negative cells mainly reflect ratio sensitivity at very small baseline errors.
}
\label{fig:bessel_j2_relative_improvement}
\vskip -0.15in
\end{figure}
The relative-improvement heatmaps (\figureautorefname{\ref{fig:bessel_j2_relative_improvement}}) provide a complementary, ratio-based view of the gains achieved by each variant over the standard QFWP. Using the metric defined earlier (\equationautorefname{\ref{eq:relative_improvement}}), where positive values indicate lower MSE than the standard baseline, both the Self-Modulating and Only Old variants exhibit strong positive improvements across most settings on the \texttt{bessel\_j2} task. In particular, for sequence lengths 16, 32, and 64, their relative-improvement values are almost uniformly close to 1, indicating highly significant gains over the standard QFWP. By contrast, the Only New variant shows much weaker and less consistent improvements, with many entries remaining near zero or even becoming negative, suggesting that new-only modulation is insufficient to reproduce the benefits of old-related modulation. A few isolated large negative values appear in the Self-Modulating heatmap; however, these cases mainly arise because the standard QFWP already attains an extremely small final test MSE in those settings, making the ratio-based metric highly sensitive to tiny absolute differences. In practice, the modulating variants in these cases also converge successfully, and their prediction curves are often visually almost indistinguishable from those of the standard model. Therefore, such isolated negative entries should not be interpreted as failures of the modulating variants, but rather read together with the raw final-test-MSE heatmaps and prediction plots. Overall, these results are consistent with the preceding analyses and further support the conclusion that old-state modulation is the dominant source of performance gain for the \texttt{bessel\_j2} task, while full self-modulation preserves this advantage across most configurations.
\begin{figure}[htbp]
\vskip -0.1in
\centering
\includegraphics[width=1\columnwidth]{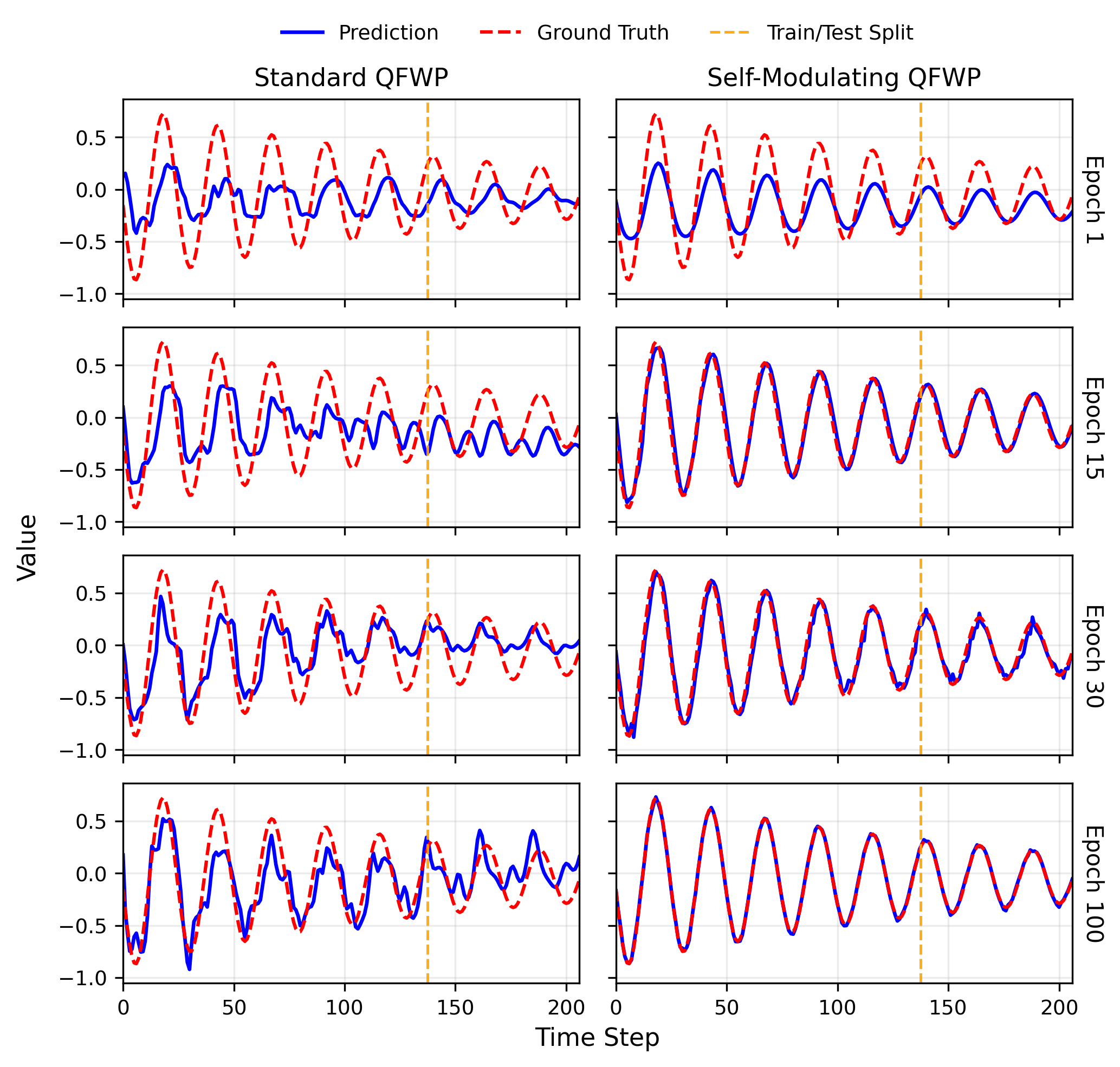}
\caption{\textbf{Prediction trajectories of Standard QFWP and Self-Modulating QFWP on the \texttt{damped\_shm} task at selected training epochs (\texttt{seq\_len}=32).} Blue solid lines denote model predictions, red dashed lines denote ground truth, and orange dashed lines indicate the train/test split.}
\label{fig:damped_shm_rollout}
\vskip -0.15in
\end{figure}

For the \texttt{damped\_shm} task with sequence length 32 (shown in \figureautorefname{\ref{fig:damped_shm_rollout}}), the Self-Modulating QFWP clearly outperforms the standard QFWP. By around epoch 15, the Self-Modulating model already produces predictions that nearly overlap with the ground truth and continues to track the damped oscillatory trajectory accurately in both the training and testing regions. In contrast, the standard QFWP still exhibits noticeable shape and amplitude mismatches even at epoch 100, indicating that it fails to fully capture the underlying temporal dynamics. Overall, this task again supports the benefit of self-modulation for sequence modeling, with an even more pronounced advantage than in \texttt{bessel\_j2}.
\begin{figure}[htbp]
\vskip -0.1in
\centering
\includegraphics[width=1\columnwidth]{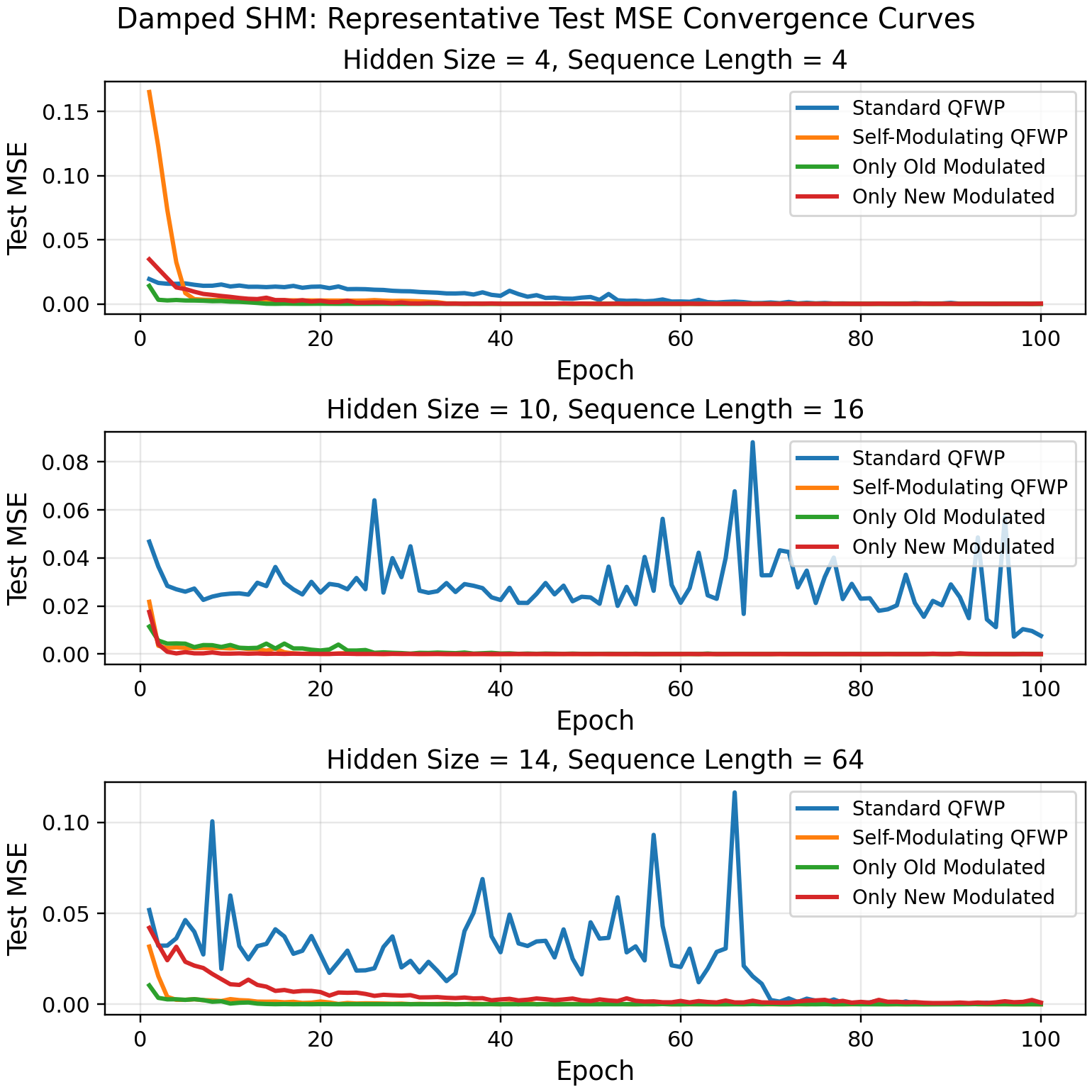}
\caption{\textbf{Representative test MSE convergence curves on the \texttt{damped\_shm} task, comparing Standard QFWP, full Self-Modulating QFWP, and its ablation variants under selected hidden sizes and sequence lengths.}}
\label{fig:damped_shm_convergence}
\vskip -0.15in
\end{figure}

The representative test MSE convergence curves for \texttt{damped\_shm} (\figureautorefname{\ref{fig:damped_shm_convergence}}) show a pattern consistent with, but even stronger than, that of \texttt{bessel\_j2}. The standard QFWP maintains substantially higher and more fluctuating test error in the medium-to-long sequence settings, whereas both the Self-Modulating QFWP and the Only Old Modulated variant rapidly converge to near-zero test MSE and remain stable thereafter. 
The Only-New variant improves over the standard model, but still lags behind the former two. This indicates that \texttt{damped\_shm} is especially sensitive to old-state modulation, which directly improves the stability of temporal memory.
\begin{figure}[htbp]
\vskip -0.1in
\centering
\includegraphics[width=1\columnwidth]{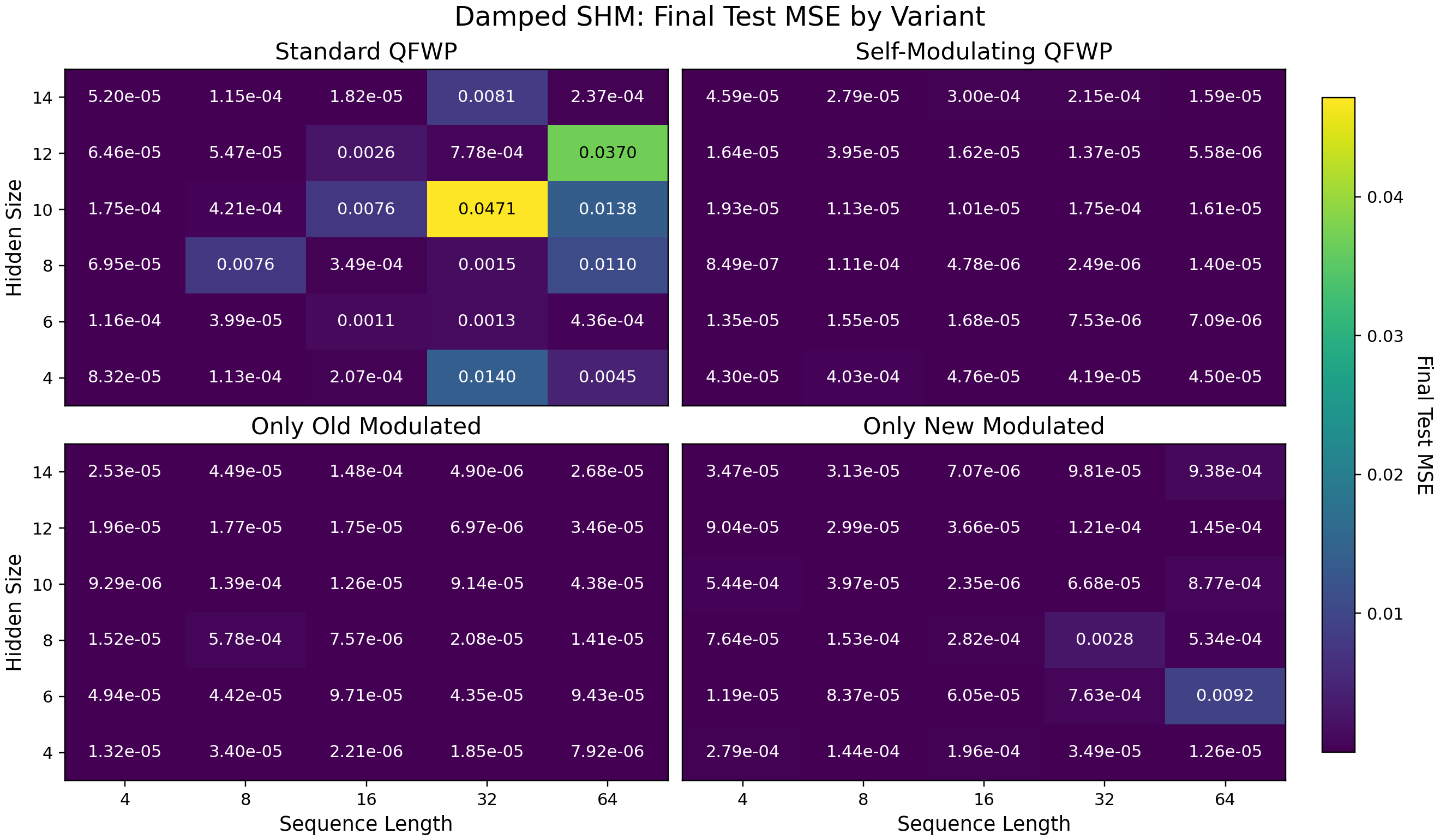}
\caption{\textbf{Final test MSE of Standard QFWP, Self-Modulating QFWP, and its ablation variants on the \texttt{damped\_shm} task across hidden sizes and sequence lengths.} Each cell reports the final test MSE for the corresponding configuration.}
\label{fig:damped_shm_variants_loss_comparison}
\vskip -0.15in
\end{figure}
The final-test-MSE heatmaps in \figureautorefname{\ref{fig:damped_shm_variants_loss_comparison}} confirm the same trend across the full configuration grid. The standard QFWP deteriorates substantially in several medium-to-long sequence settings, while the Self-Modulating QFWP remains low-error and stable across hidden sizes and sequence lengths. The Only-Old variant is again among the most consistent performers, whereas the Only-New variant provides partial improvement but remains less robust. This reinforces the conclusion that the dominant gain in \texttt{damped\_shm} comes from old-related modulation.
\begin{figure}[htbp]
\vskip -0.1in
\centering
\includegraphics[width=1\columnwidth]{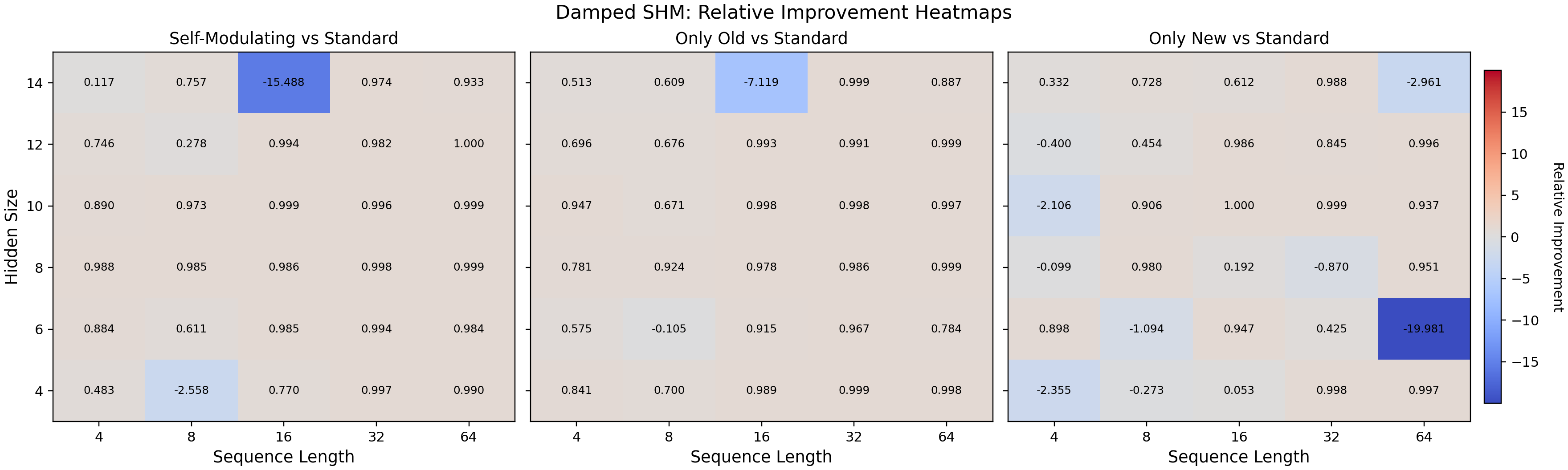}
\caption{\textbf{Relative improvement over Standard QFWP on the \texttt{damped\_shm} task for Self-Modulating QFWP and its ablation variants across hidden sizes and sequence lengths.} Positive values indicate improvement over the standard baseline, while negative values indicate degradation.}
\label{fig:damped_shm_relative_improvement}
\vskip -0.15in
\end{figure}

The relative-improvement heatmaps in \figureautorefname{\ref{fig:damped_shm_relative_improvement}} further show that both the Self-Modulating and Only-Old variants achieve strong positive gains over the standard QFWP across most settings, with many values approaching 1 in the medium-to-long sequence regime. The Only-New variant is less consistent and still contains several near-zero or negative regions. As in the \texttt{bessel\_j2} case, isolated negative values should be interpreted together with the raw MSE heatmaps, since the ratio-based metric can amplify tiny absolute differences when the baseline error is already extremely small.
\begin{figure}[htbp]
\vskip -0.1in
\centering
\includegraphics[width=1\columnwidth]{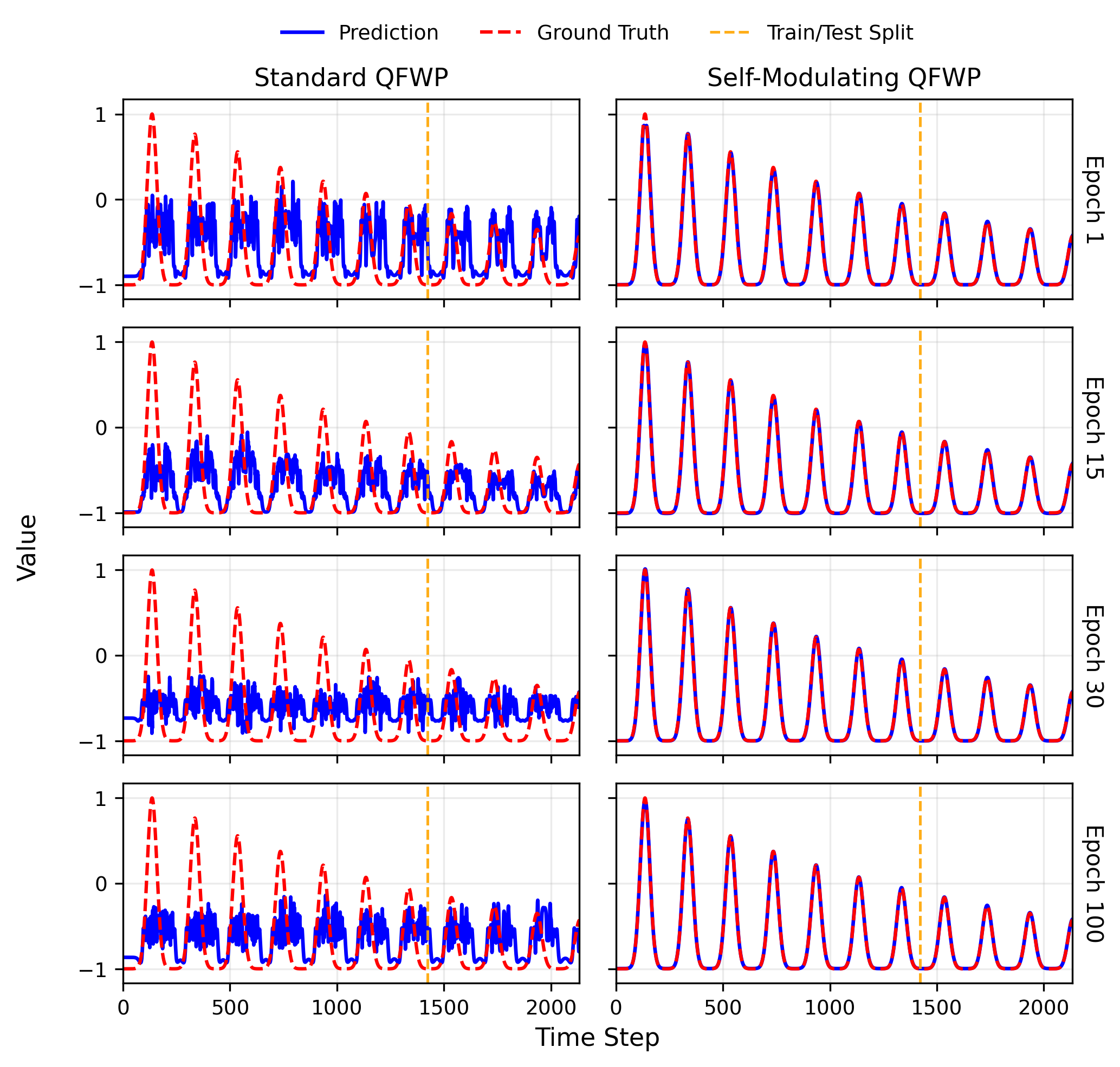}
\caption{\textbf{Prediction trajectories of Standard QFWP and Self-Modulating QFWP on the \texttt{delayed\_quantum\_control} task at selected training epochs (\texttt{seq\_len}=64).} Blue solid lines denote model predictions, red dashed lines denote ground truth, and orange dashed lines indicate the train/test split.}
\label{fig:delayed_quantum_control_rollout}
\vskip -0.1in
\end{figure}

For the \texttt{delayed\_quantum\_control} task with sequence length 64 (shown in \figureautorefname{\ref{fig:delayed_quantum_control_rollout}}), the contrast between the two models is nearly binary. The Self-Modulating QFWP reproduces the pulse-like decaying target dynamics almost perfectly from very early epochs and remains highly accurate in both the training and testing regions. In contrast, the standard QFWP fails to learn the correct temporal structure throughout training and still exhibits clearly mismatched low-amplitude oscillatory predictions even at epoch 100. This suggests that, in this task, self-modulation does not merely improve convergence speed, but is crucial for successfully learning the delayed temporal dynamics.
\begin{figure}[htbp]
\vskip -0.1in
\centering
\includegraphics[width=1\columnwidth]{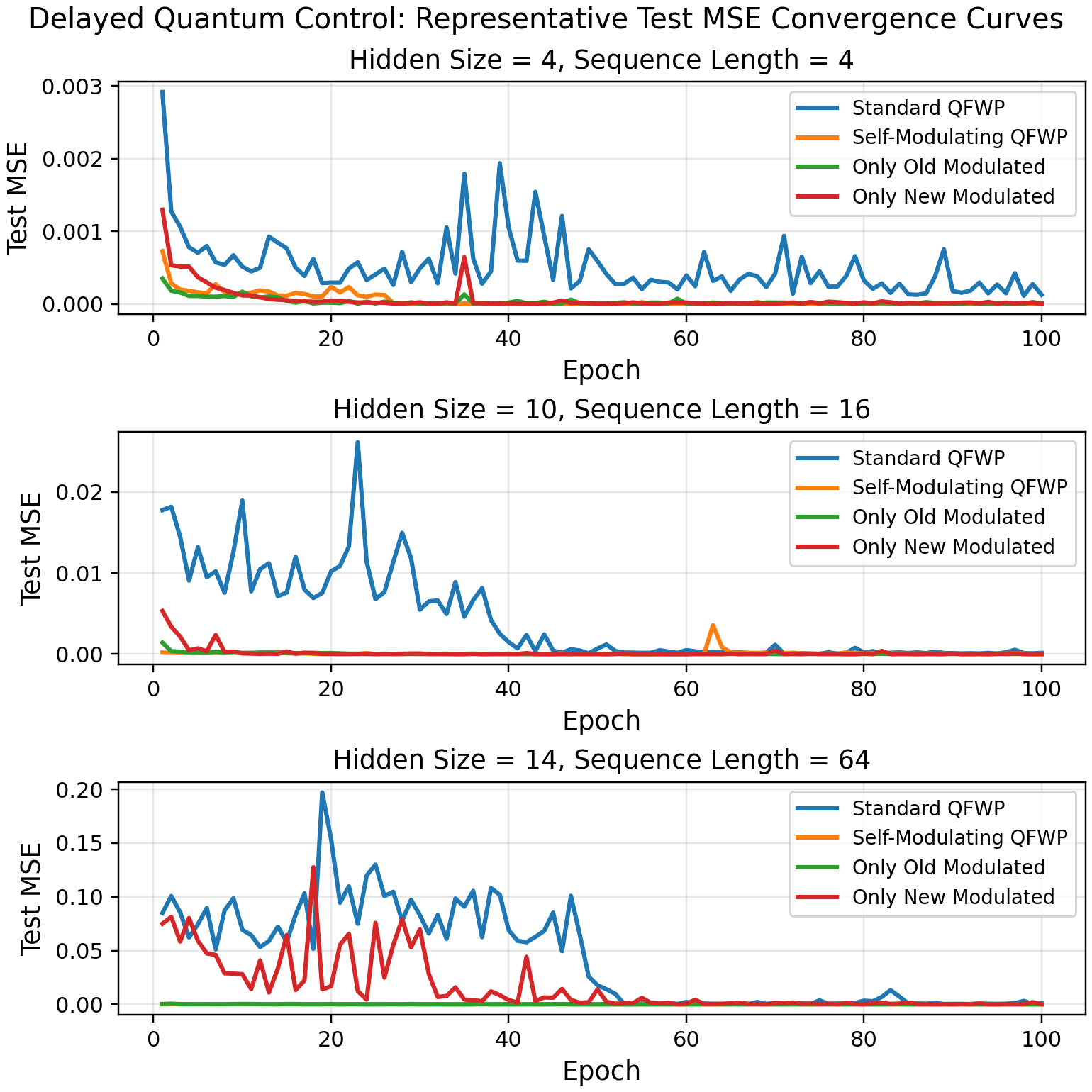}
\caption{\textbf{Representative test MSE convergence curves on the \texttt{delayed\_quantum\_control} task, comparing Standard QFWP, full Self-Modulating QFWP, and its ablation variants under selected hidden sizes and sequence lengths.}}
\label{fig:delayed_quantum_control_convergence}
\vskip -0.1in
\end{figure}

The convergence curves in \figureautorefname{\ref{fig:delayed_quantum_control_convergence}} show that old-related modulation is highly effective for this task. Both the Self-Modulating and Only-Old variants remain near zero and highly stable from early training, whereas the standard QFWP exhibits substantially larger errors and stronger fluctuations in medium-to-long sequence settings. The Only-New variant also improves over the standard model, but remains less stable than the Self-Modulating and Only-Old variants. These results indicate that old-state modulation almost eliminates the original optimization difficulty in \texttt{delayed\_quantum\_control}.
\begin{figure}[htbp]
\vskip -0.1in
\centering
\includegraphics[width=1\columnwidth]{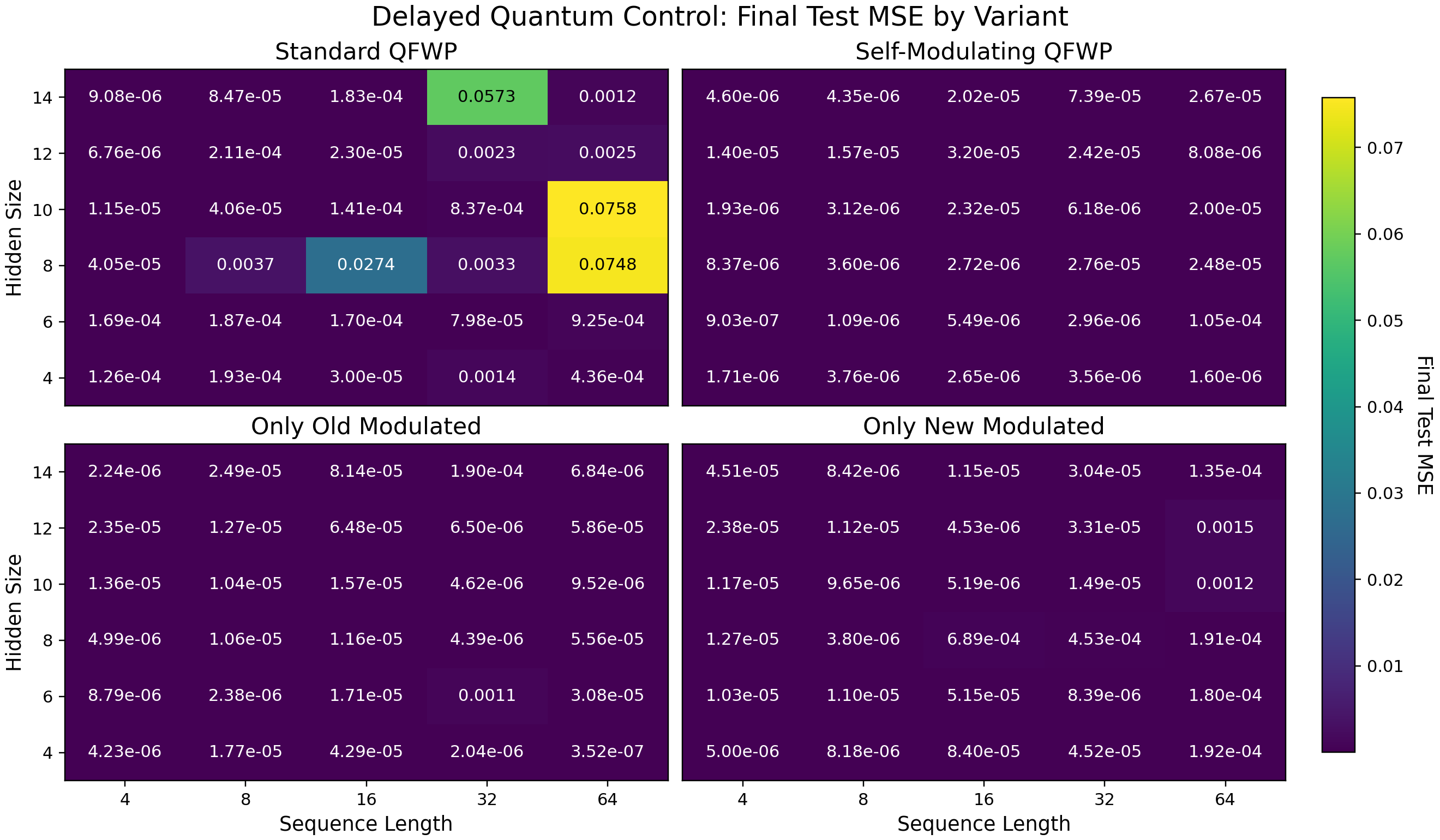}
\caption{\textbf{Final test MSE of Standard QFWP, Self-Modulating QFWP, and its ablation variants on the \texttt{delayed\_quantum\_control} task across hidden sizes and sequence lengths.} Each cell reports the final test MSE for the corresponding configuration.}
\label{fig:delayed_quantum_control_loss_comparison}
\vskip -0.1in
\end{figure}

The final-test-MSE heatmaps for \texttt{delayed\_quantum\_control} (\figureautorefname{\ref{fig:delayed_quantum_control_loss_comparison}}) reveal an even sharper separation than in the previous tasks. The standard QFWP exhibits several clearly failed medium-to-long sequence settings with markedly elevated final test MSE, indicating limited ability to model the delayed temporal structure. In contrast, the Self-Modulating QFWP maintains uniformly low and stable error across almost the entire hidden-size and sequence-length grid. 
The Only-Old variant performs comparably well and shows similarly consistent low-error behavior, while the Only-New variant, although better than the standard model, still retains a few higher-error configurations in harder settings.
Overall, these results again suggest that the dominant gain in \texttt{delayed\_quantum\_control} comes from old-related modulation, while full self-modulation preserves this advantage across the configuration space.
\begin{figure}[htbp]
\vskip -0.1in
\centering
\includegraphics[width=1\columnwidth]{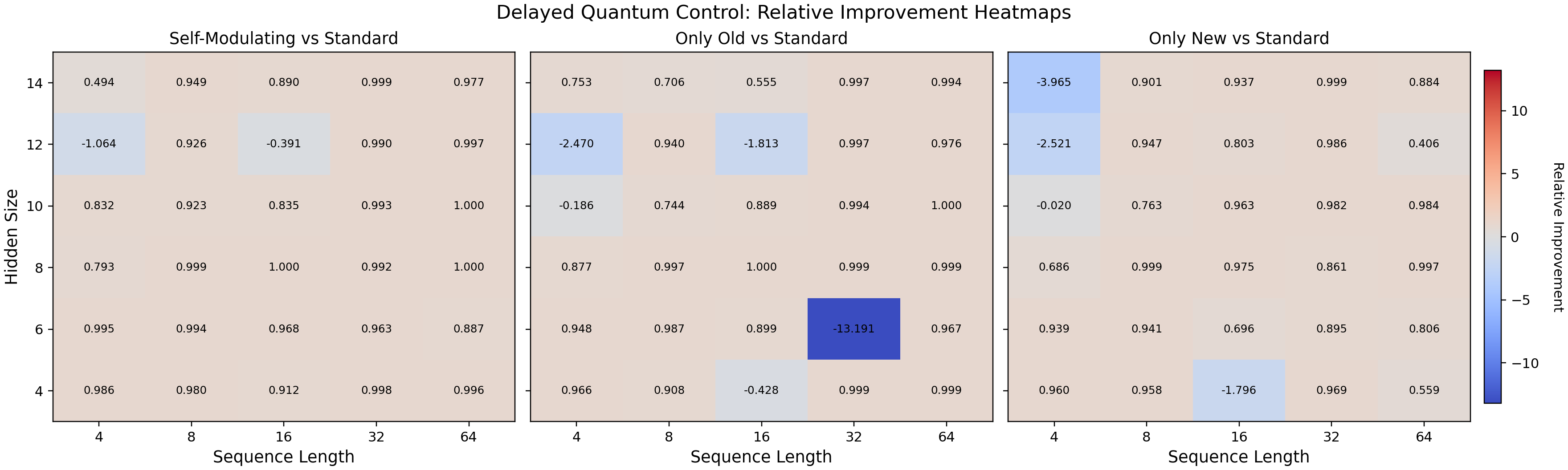}
\caption{\textbf{Relative improvement over Standard QFWP on the \texttt{delayed\_quantum\_control} task for Self-Modulating QFWP and its ablation variants across hidden sizes and sequence lengths.} Positive values indicate improvement over the standard baseline, while negative values indicate degradation.}
\label{fig:delayed_quantum_control_relative_improvement}
\vskip -0.1in
\end{figure}

\begin{figure}[htbp]
\vskip -0.1in
\centering
\includegraphics[width=1\columnwidth]{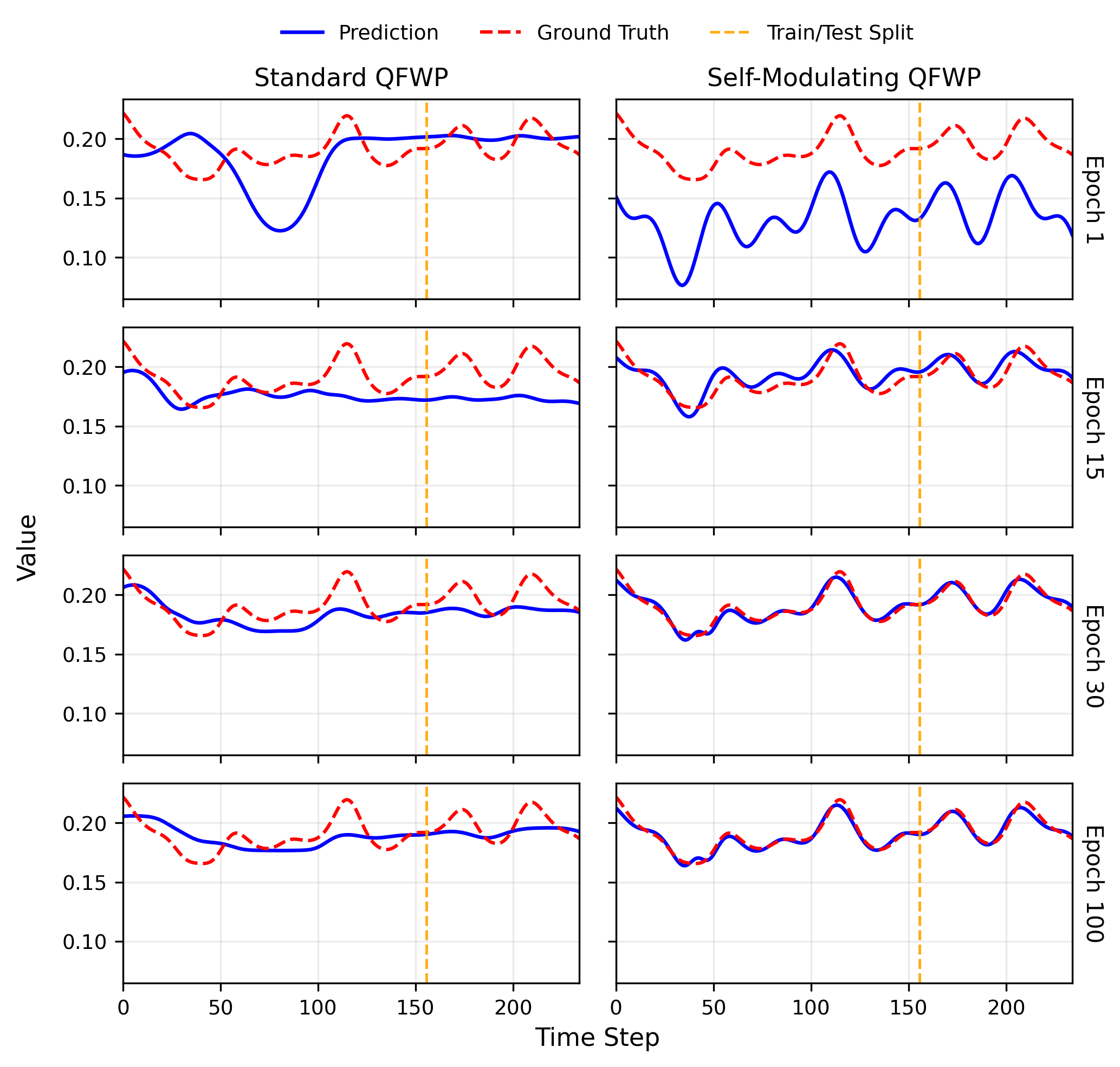}
\caption{\textbf{Prediction trajectories of Standard QFWP and Self-Modulating QFWP on the \texttt{narma\_10} task at selected training epochs (\texttt{seq\_len}=64).} Blue solid lines denote model predictions, red dashed lines denote ground truth, and orange dashed lines indicate the train/test split.}
\label{fig:narma_10_rollout}
\vskip -0.1in
\end{figure}

\begin{figure}[htbp]
\vskip -0.1in
\centering
\includegraphics[width=1\columnwidth]{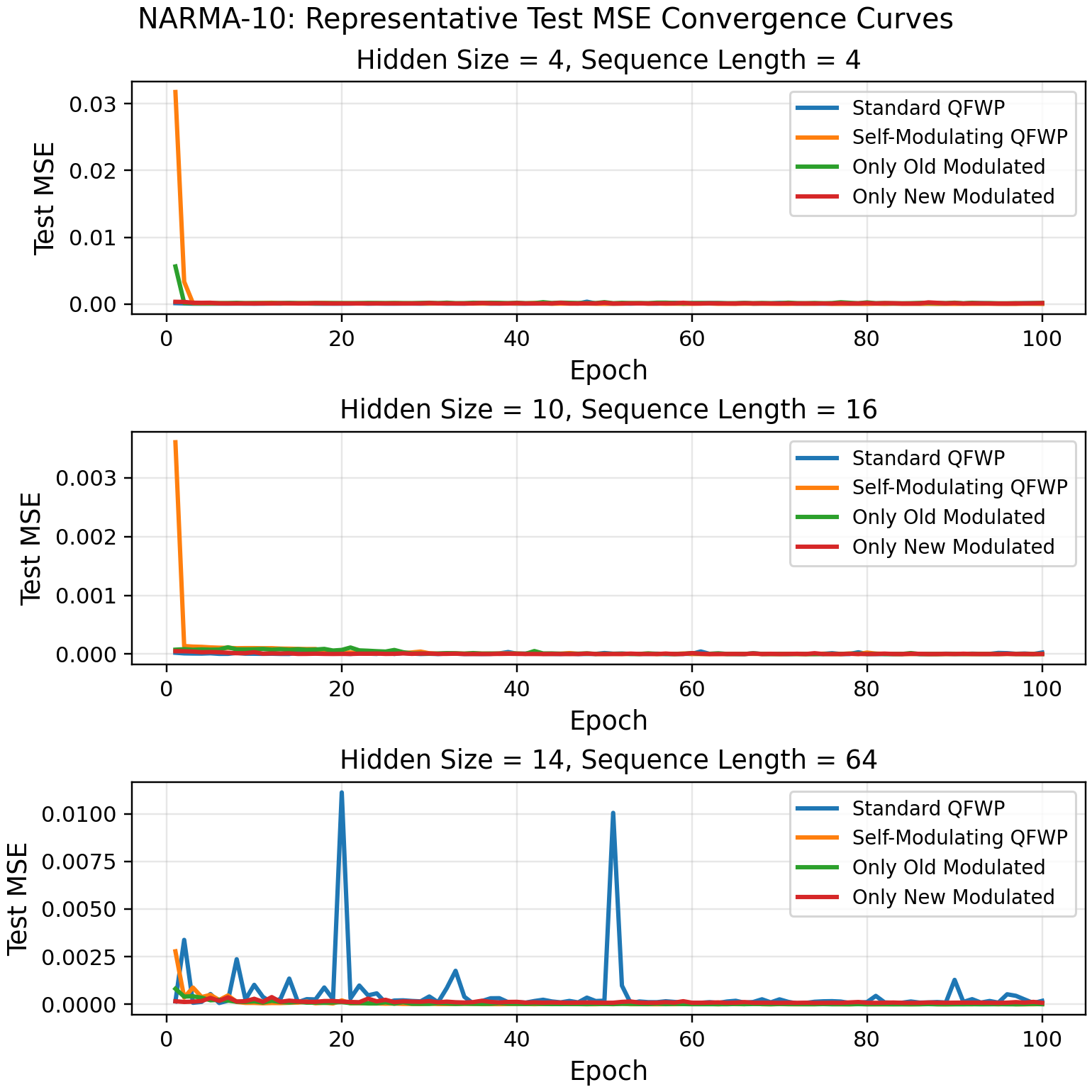}
\caption{\textbf{Representative test MSE convergence curves on the \texttt{narma\_10} task, comparing Standard QFWP, full Self-Modulating QFWP, and its ablation variants under selected hidden sizes and sequence lengths.}}
\label{fig:narma_10_convergence}
\vskip -0.1in
\end{figure}

\begin{figure}[htbp]
\vskip -0.1in
\centering
\includegraphics[width=1\columnwidth]{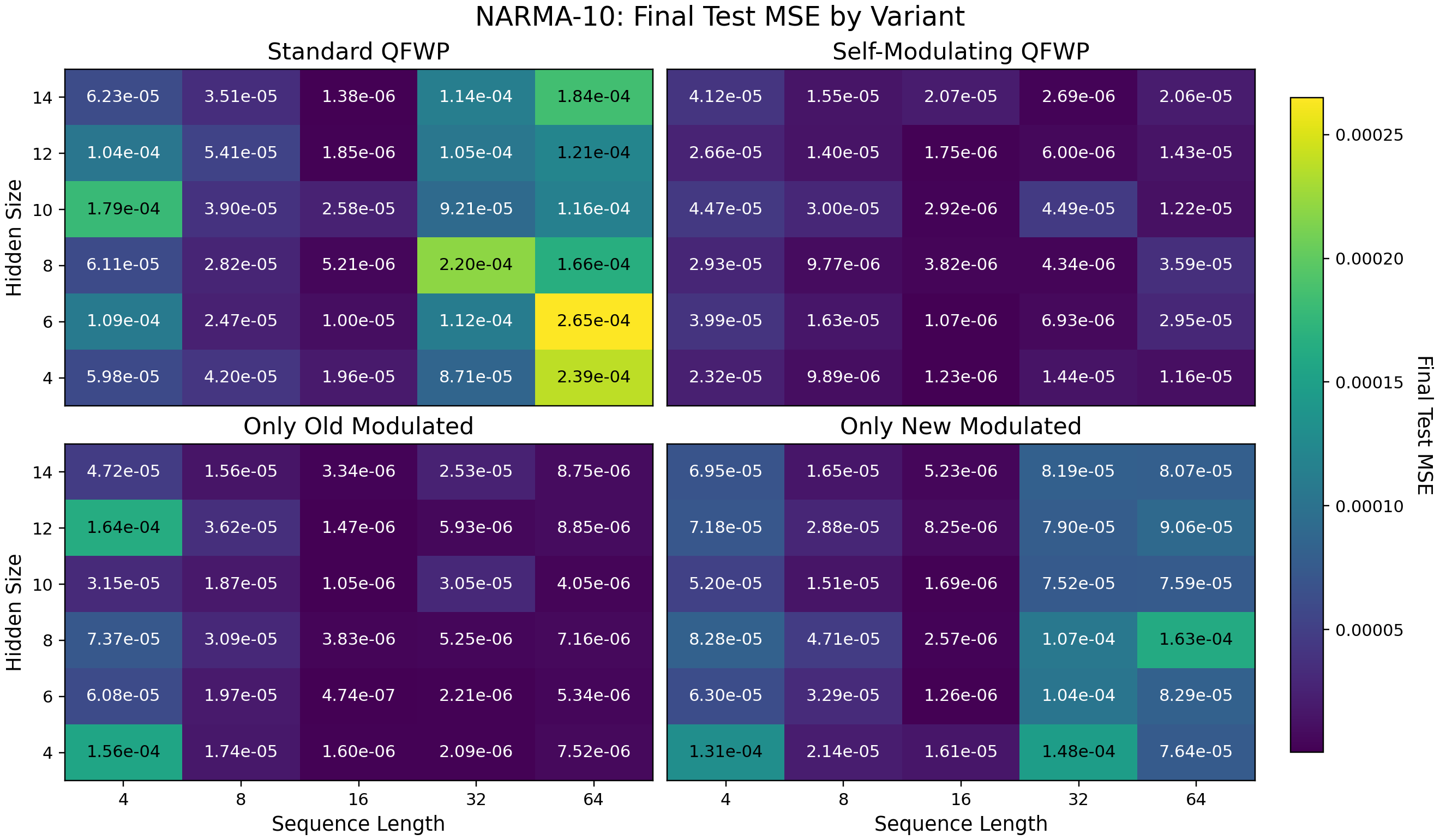}
\caption{\textbf{Final test MSE of Standard QFWP, Self-Modulating QFWP, and its ablation variants on the \texttt{narma\_10} task across hidden sizes and sequence lengths.} Each cell reports the final test MSE for the corresponding configuration.}
\label{fig:narma_10_loss_comparison}
\vskip -0.1in
\end{figure}

\begin{figure}[htbp]
\vskip -0.1in
\centering
\includegraphics[width=1\columnwidth]{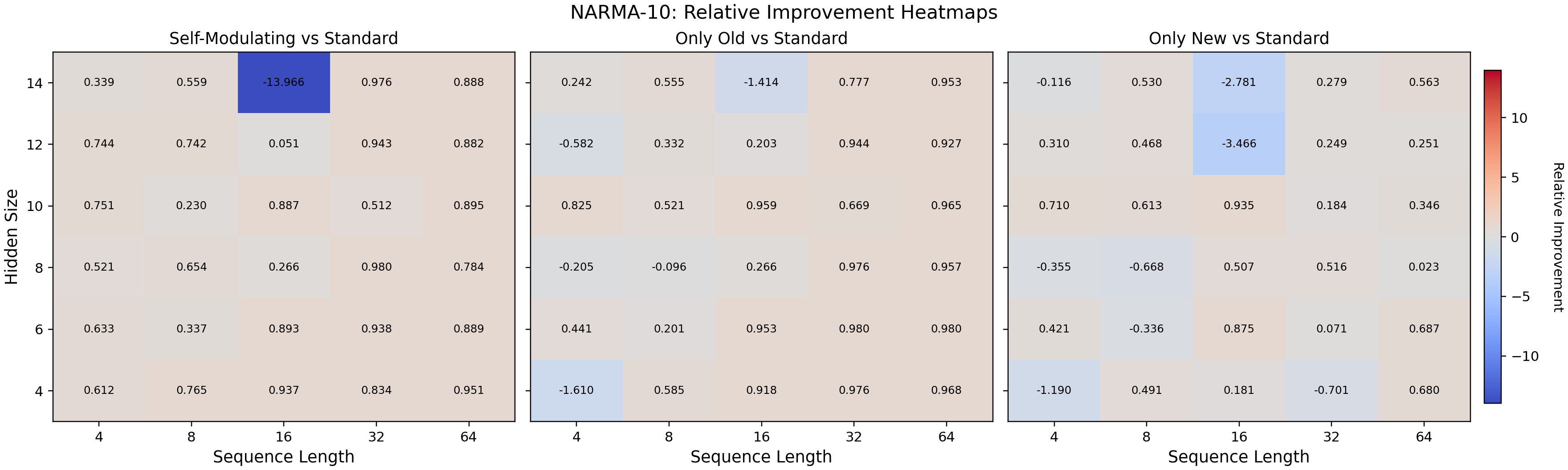}
\caption{\textbf{Relative improvement over Standard QFWP on the \texttt{narma\_10} task for Self-Modulating QFWP and its ablation variants across hidden sizes and sequence lengths.} Positive values indicate improvement over the standard baseline, while negative values indicate degradation.}
\label{fig:narma_10_relative_improvement}
\vskip -0.1in
\end{figure}

The relative-improvement heatmaps in \figureautorefname{\ref{fig:delayed_quantum_control_relative_improvement}} are consistent with the raw MSE results. Both the Self-Modulating and Only-Old variants achieve strong positive gains over the standard QFWP across most settings, especially in the longer-sequence regime. The Only-New variant also improves over the standard baseline, but its gains are generally less uniform. Overall, these results again support the conclusion that the dominant gain in \texttt{delayed\_quantum\_control} comes from old-related modulation, while full self-modulation preserves this advantage across the configuration space.
\begin{figure}[htbp]
\vskip -0.1in
\centering
\includegraphics[width=1\columnwidth]{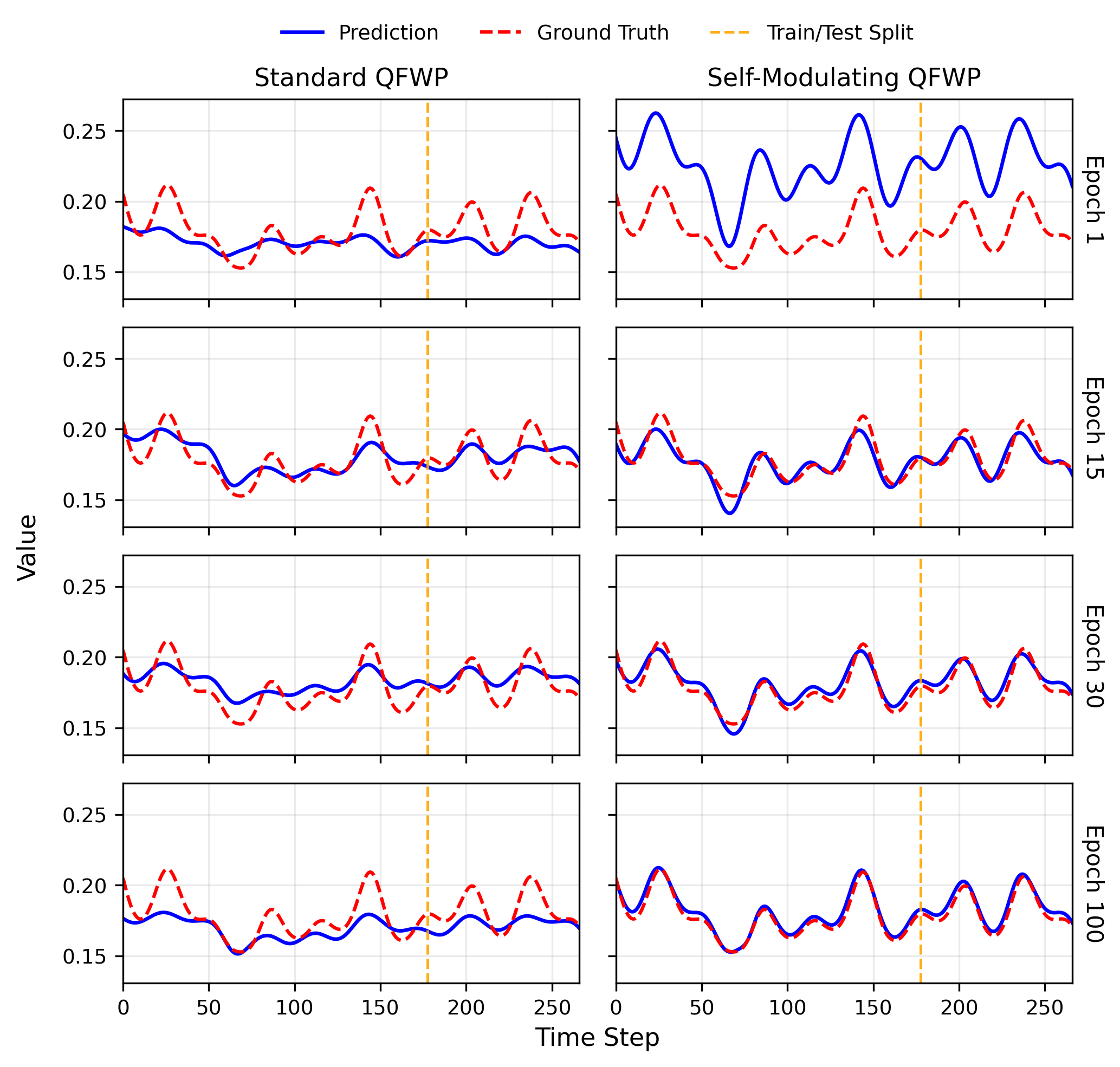}
\caption{\textbf{Prediction trajectories of Standard QFWP and Self-Modulating QFWP on the \texttt{narma\_5} task at selected training epochs (\texttt{seq\_len}=32).} Blue solid lines denote model predictions, red dashed lines denote ground truth, and orange dashed lines indicate the train/test split.}
\label{fig:narma_5_rollout}
\vskip -0.1in
\end{figure}
\begin{figure}[htbp]
\vskip -0.1in
\centering
\includegraphics[width=1\columnwidth]{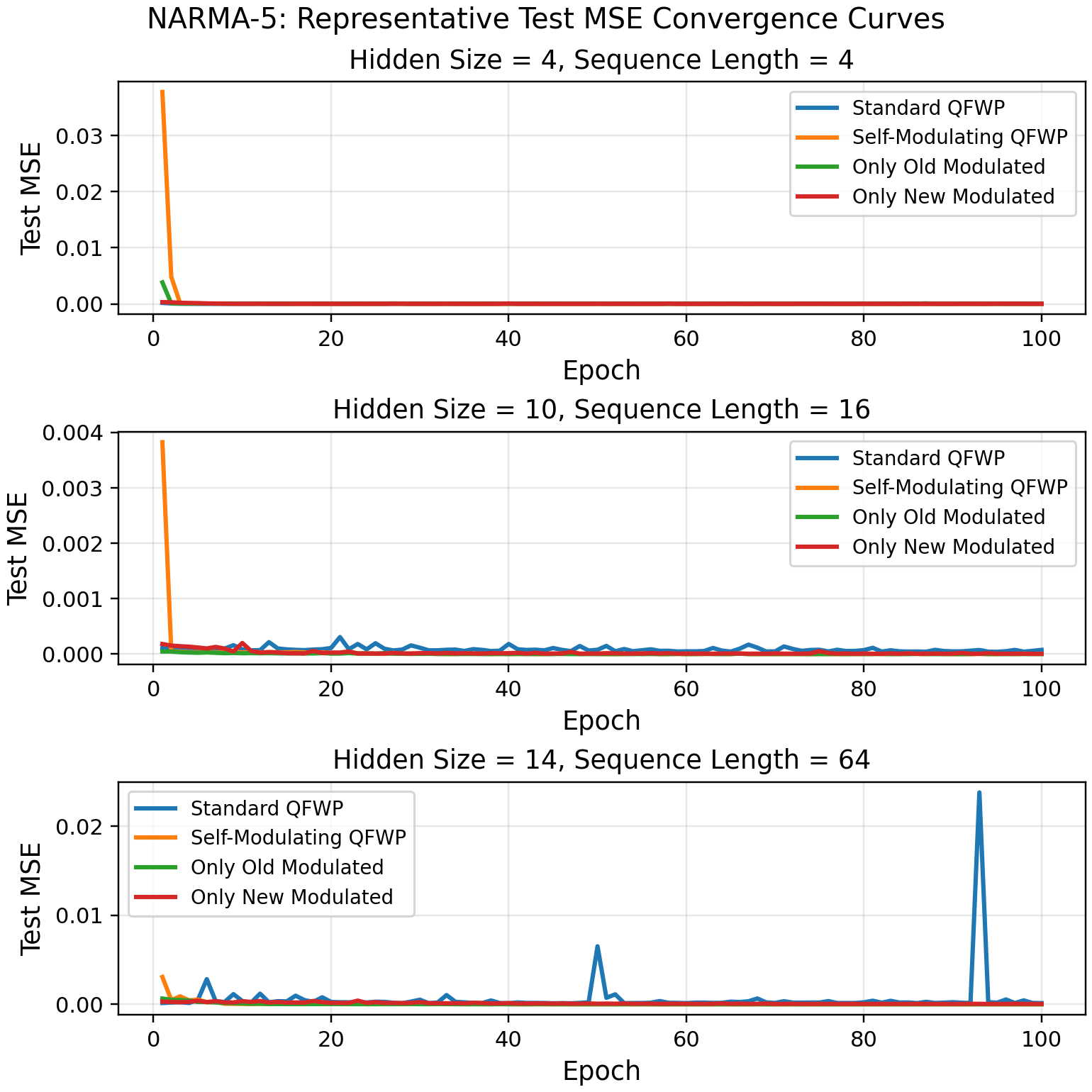}
\caption{\textbf{Representative test MSE convergence curves on the \texttt{narma\_5} task task, comparing Standard QFWP, full Self-Modulating QFWP, and its ablation variants under selected hidden sizes and sequence lengths.}
}
\label{fig:narma_5_convergence}
\vskip -0.1in
\end{figure}
\begin{figure}[htbp]
\vskip -0.1in
\centering
\includegraphics[width=1\columnwidth]{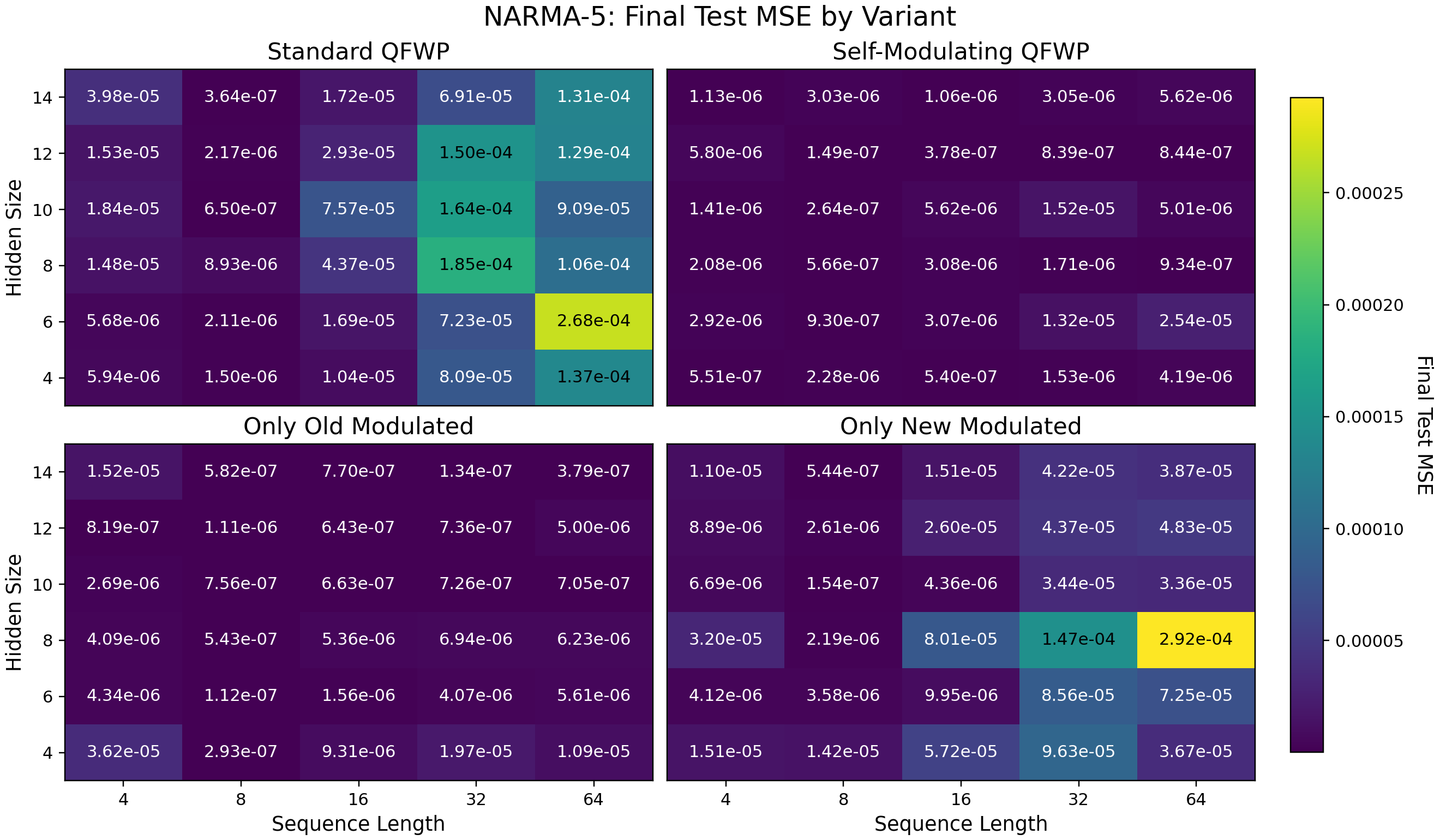}
\caption{\textbf{Final test MSE of Standard QFWP, Self-Modulating QFWP, and its ablation variants on the \texttt{narma\_5} task across hidden sizes and sequence lengths.} Each cell reports the final test MSE for the corresponding configuration.}
\label{fig:narma_5_loss_comparison}
\end{figure}
\begin{figure}[htbp]
\vskip -0.1in
\centering
\includegraphics[width=1\columnwidth]{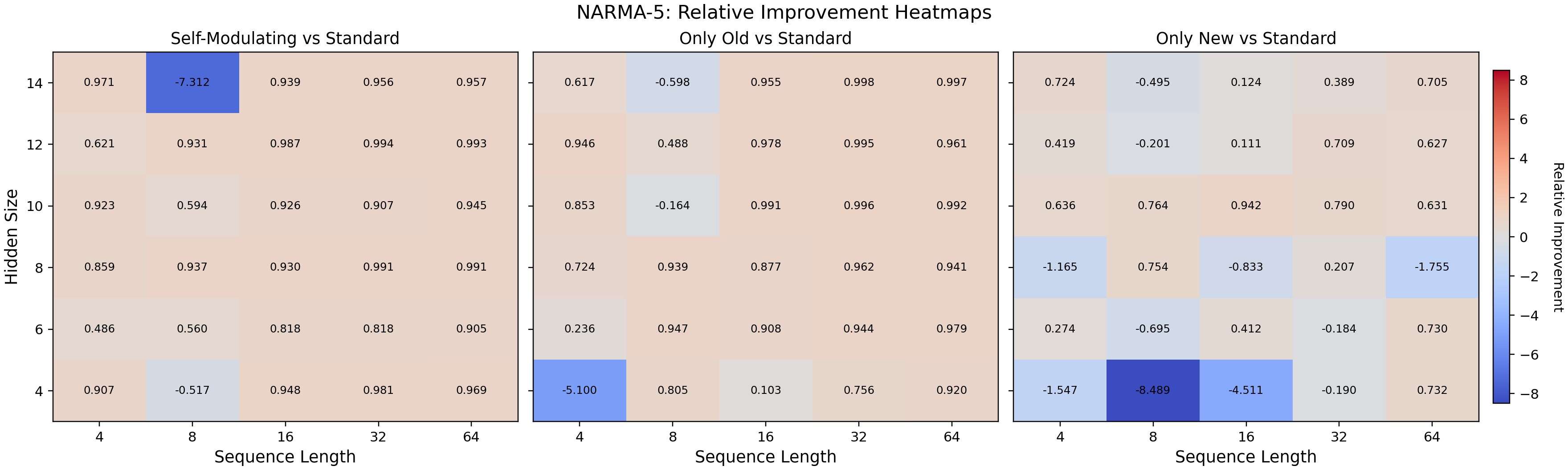}
\caption{\textbf{Relative improvement over Standard QFWP on the \texttt{narma\_5} task for Self-Modulating QFWP and its ablation variants across hidden sizes and sequence lengths.} 
}
\label{fig:narma_5_relative_improvement}
\vskip -0.1in
\end{figure}
We next examine the NARMA family---\texttt{narma\_5} (sequence length 32) and \texttt{narma\_10} (sequence length 64)---as progressively harder autoregressive memory tasks.
The same overall ordering holds in both: Self-Modulating $\approx$ Only-Old $>$ Only-New $>$ standard QFWP, visible across rollouts (\figureautorefname{\ref{fig:narma_5_rollout}}, \figureautorefname{\ref{fig:narma_10_rollout}}), convergence curves (\figureautorefname{\ref{fig:narma_5_convergence}}, \figureautorefname{\ref{fig:narma_10_convergence}}), final-MSE heatmaps (\figureautorefname{\ref{fig:narma_5_loss_comparison}}, \figureautorefname{\ref{fig:narma_10_loss_comparison}}), and relative-improvement heatmaps (\figureautorefname{\ref{fig:narma_5_relative_improvement}}, \figureautorefname{\ref{fig:narma_10_relative_improvement}}).

For \texttt{narma\_5} the standard QFWP captures the coarse trend but smooths out local fluctuations and peak--valley structure; in convergence it eventually reaches low error but with greater instability and occasional spikes in the harder settings, while Self-Modulating and Only-Old stay smooth and low.
The same ordering becomes more pronounced for \texttt{narma\_10}: the standard model is more clearly underfit in the rollouts, fluctuates more sharply in the convergence curves, and degrades more visibly in the heatmaps, while full Self-Modulating and Only-Old variants converge smoothly and remain near-zero loss.
The widening gap from \texttt{narma\_5} to \texttt{narma\_10} indicates that old-related modulation becomes increasingly important as autoregressive memory demand grows.


The convergence curves show that all variants can eventually reach relatively low test error, but the standard QFWP exhibits greater instability and occasional spikes in harder settings, whereas the Self-Modulating and Only-Old variants remain more stable overall (\figureautorefname{\ref{fig:narma_5_convergence}}). The final-test-MSE and relative-improvement heatmaps further confirm this ranking: Self-Modulating and Only-Old achieve the most consistent gains, Only-New provides intermediate improvement, and the standard QFWP is the least robust, especially at longer sequence lengths (\figureautorefname{\ref{fig:narma_5_loss_comparison}} and \figureautorefname{\ref{fig:narma_5_relative_improvement}}).

The same ordering becomes more pronounced for \texttt{narma\_10} with sequence length 64. In the epoch-wise prediction comparison, the standard QFWP remains more underfit and less responsive to local variations, whereas the Self-Modulating variant tracks the target sequence more faithfully from mid training onward (\figureautorefname{\ref{fig:narma_10_rollout}}). This stronger gap is also visible in the convergence curves, where the standard model exhibits larger fluctuations and sharper spikes in harder settings, while the Self-Modulating and Only-Old variants converge more smoothly and maintain lower test MSE (\figureautorefname{\ref{fig:narma_10_convergence}}). 

The final-test-MSE and relative-improvement heatmaps again identify Self-Modulating and Only-Old as the most reliable configurations across hidden sizes and sequence lengths, with Only-New remaining intermediate and the standard QFWP showing the most obvious degradation (\figureautorefname{\ref{fig:narma_10_loss_comparison}} and \figureautorefname{\ref{fig:narma_10_relative_improvement}}). Compared with \texttt{narma\_5}, the stronger separation in \texttt{narma\_10} suggests that old-related modulation becomes increasingly important as the autoregressive memory demand increases.

\begin{figure*}[htbp]
\vskip -0.1in
\centering
\includegraphics[width=1\textwidth]{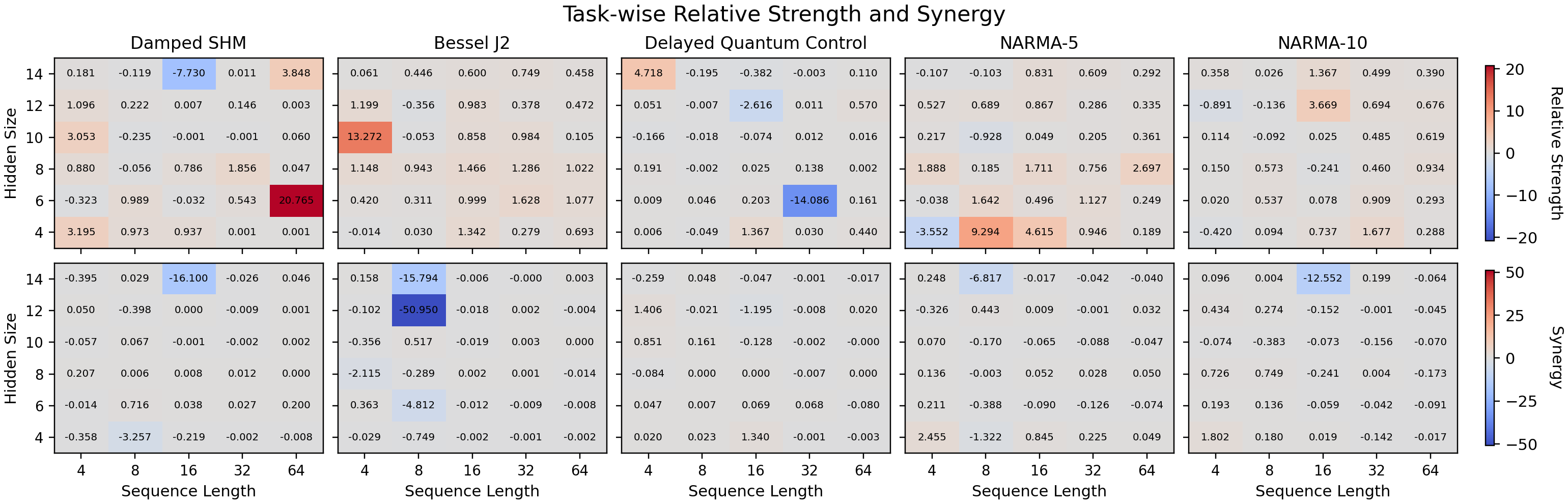}
\caption{\textbf{Task-wise summary heatmaps of relative strength and synergy across hidden sizes and sequence lengths for the five benchmark tasks.} Numeric annotations report the corresponding summary scores for each configuration.}
\label{fig:all_relative_strength_and_synergy}
\vskip -0.1in
\end{figure*}
\begin{figure}[htbp]
\vskip -0.1in
\centering
\includegraphics[width=1\columnwidth]{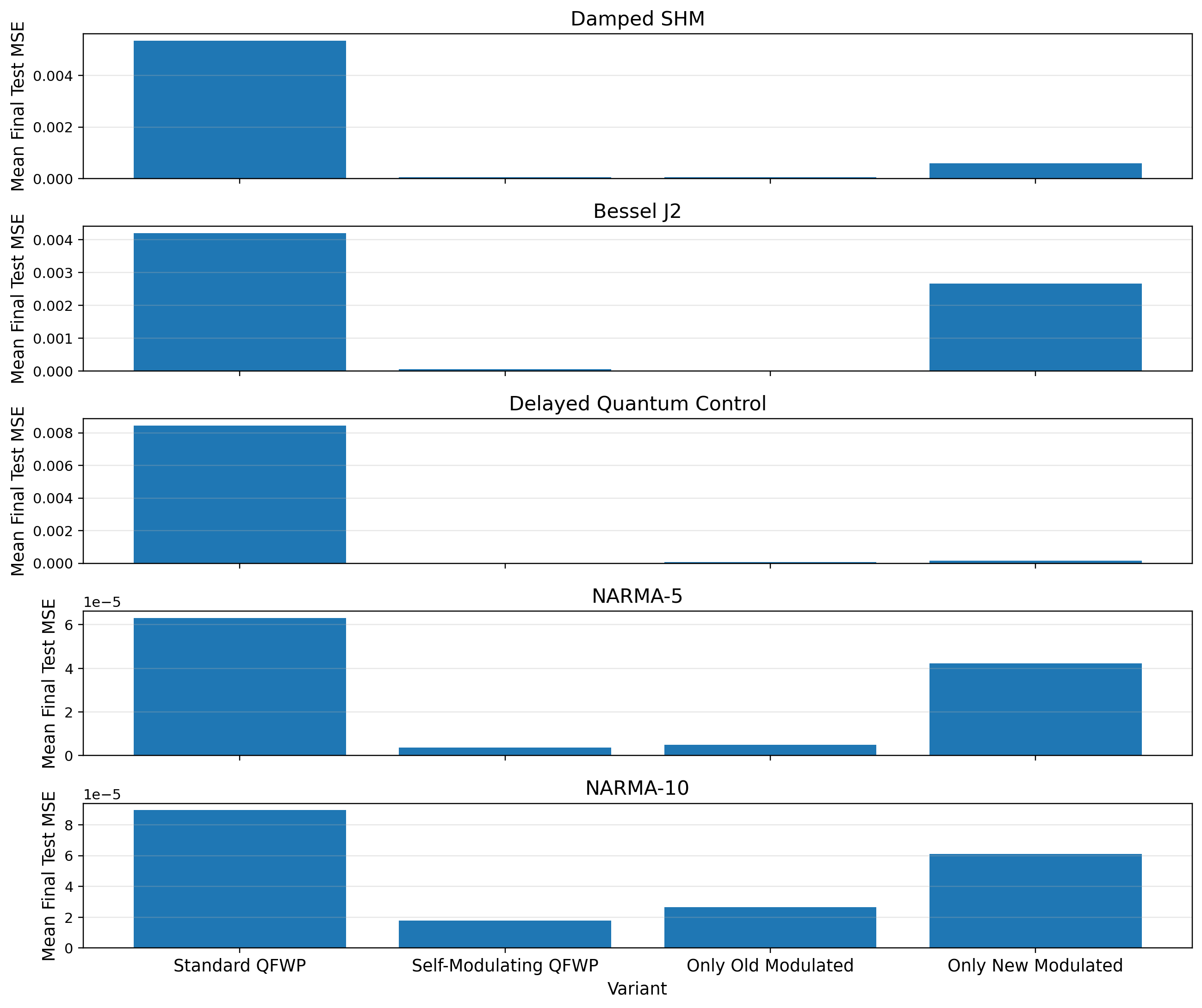}
\caption{\textbf{Task-wise mean final test MSE of Standard QFWP, Self-Modulating QFWP, and its ablation variants, providing an aggregate comparison of overall predictive performance across configurations. Lower values indicate better performance.}
}
\label{fig:all_final_aggregated_loss}
\vskip -0.15in
\end{figure}

The task-wise mean final test MSE (\figureautorefname{\ref{fig:all_final_aggregated_loss}}) confirms the same ranking at the aggregate level across all five benchmarks: standard QFWP is weakest, Only-New gives partial improvement, and full Self-Modulating and the Only-Old variant are strongest.
The advantage is most dramatic on the oscillatory and delayed-control tasks, where full Self-Modulating and Only-Old reduce mean MSE by orders of magnitude; on NARMA absolute errors are smaller but the ordering is unchanged.
The gain from self-modulation is therefore broadly task-independent.

The Relative Strength and Synergy heatmaps (\figureautorefname{\ref{fig:all_relative_strength_and_synergy}}, \equationautorefname{\ref{eq:relative_strength}}, \equationautorefname{\ref{eq:synergy}}) interpret these gains mechanistically.
Recall that positive Relative Strength means old-state modulation contributes more than new-state modulation, while positive Synergy means the full model exceeds the better single-sided variant.
Across nearly all tasks and configurations, Relative Strength is predominantly positive---often strongly so---confirming that the gain is dominated by old-related modulation.
By contrast, Synergy values are frequently near zero or negative, especially on \texttt{bessel\_j2} and \texttt{damped\_shm}, indicating that the full Self-Modulating model rarely benefits from constructive old/new interaction.
Its performance is largely explained by the dominant old-state effect, with new modulation providing only limited or task-dependent additional benefit.
NARMA shows a milder, mixed pattern but the same direction.
Together with the per-task results, these aggregate views support the conclusion that old-state modulation is the primary driver of improvement, and full self-modulation mainly serves as a robust wrapper preserving this benefit across diverse tasks.

\section{Theoretical Discussion}

Here we provide an explanation of why the \emph{Only-Old} model can achieve performance comparable to the full Self-Modulating QFWP.

\paragraph{Basic Notation.}
Consider a single layer-qubit coordinate \(j=(k,q)\). Define \(\theta_{t,j}=[\Theta_t]_j\), \(d_{t,j}=[\Delta_t]_j\), \(c_{t,j}=[M_t^{\mathrm{new}}]_j\), and \(a_{t,j}=[M_t^{\mathrm{old}}]_j\), where \(d_{t,j}\) is the raw update, \(c_{t,j}\) is the new-parameter modulation, and \(a_{t,j}\) is the old-parameter modulation. The original QFWP is \(\theta_{t,j}^{\mathrm{orig}}=\theta_{t-1,j}^{\mathrm{orig}}+d_{t,j}\). The Only-New variant is \(\theta_{t,j}^{\mathrm{new}}=\theta_{t-1,j}^{\mathrm{new}}+c_{t,j}d_{t,j}\). The Only-Old variant is \(\theta_{t,j}^{\mathrm{old}}=a_{t,j}\theta_{t-1,j}^{\mathrm{old}}+d_{t,j}\). The full Self-Modulating QFWP is \(\theta_{t,j}^{\mathrm{full}}=a_{t,j}\theta_{t-1,j}^{\mathrm{full}}+c_{t,j}d_{t,j}\). We define the Old-modulated temporal memory kernel as \(K_{s\to t,j}^{\mathrm{old}}=\prod_{u=s+1}^{t}a_{u,j}\), with the empty product defined as \(1\).

\paragraph{Dynamics of Updates.}
Assuming \(\Theta_0=0\), the original QFWP unrolls to \(\theta_{t,j}^{\mathrm{orig}}=\sum_{s=1}^{t}d_{s,j}\), meaning that all past updates are accumulated with equal weight. The Only-New variant unrolls to \(\theta_{t,j}^{\mathrm{new}}=\sum_{s=1}^{t}c_{s,j}d_{s,j}\), meaning that each update is rescaled when it is written, but is not dynamically reweighted afterward. The Only-Old variant unrolls to \(\theta_{t,j}^{\mathrm{old}}=\sum_{s=1}^{t}d_{s,j}\prod_{u=s+1}^{t}a_{u,j}=\sum_{s=1}^{t}d_{s,j}K_{s\to t,j}^{\mathrm{old}}\), meaning that each past update is weighted by a temporal memory kernel. The full model unrolls to \(\theta_{t,j}^{\mathrm{full}}=\sum_{s=1}^{t}c_{s,j}d_{s,j}\prod_{u=s+1}^{t}a_{u,j}=\sum_{s=1}^{t}c_{s,j}d_{s,j}K_{s\to t,j}^{\mathrm{old}}\). Thus, both the full and Only-Old models share the same temporal memory kernel \(K_{s\to t,j}^{\mathrm{old}}\); the full model only adds the instantaneous write multiplier \(c_{s,j}\).

\paragraph{Difference Between Full and Only-Old.}
With \(d_{t,j}\) and \(a_{t,j}\) fixed, only \(c_{t,j}\) differs; this is a structural, not parameter-wise, comparison. Define \(e_{t,j}=\theta_{t,j}^{\mathrm{full}}-\theta_{t,j}^{\mathrm{old}}\). Subtracting the full and Only-Old recurrences gives $e_{t,j} = a_{t,j}e_{t-1,j}+(c_{t,j}-1)d_{t,j}$.
The first term \(a_{t,j}e_{t-1,j}\) only propagates the discrepancy already present at the previous time step, while the second term \((c_{t,j}-1)d_{t,j}\) is the newly injected discrepancy caused by New modulation. Therefore, if the raw update \(d_{t,j}\) is moderate, or if the Old-modulated memory dynamics attenuate past discrepancies, the full and Only-Old models can remain close even when \(c_{t,j}\neq 1\). In matrix form, with \(E_t=\Theta_t^{\mathrm{full}}-\Theta_t^{\mathrm{old}}\), the same relation is
$
E_t
=
M_t^{\mathrm{old}}\odot E_{t-1}
+
(M_t^{\mathrm{new}}-\mathbf{1})\odot \Delta_t$.

\paragraph{Role of New Modulation.}
There is also a structural reason why New modulation may be less essential than Old modulation. The raw update has the outer-product form \(\Delta_t=\ell_t r_t^\top\), and the New modulation matrix has the outer-product form \(M_t^{\mathrm{new}}=m_t^{\mathrm{new},L}(m_t^{\mathrm{new},Q})^\top\). Their element-wise product satisfies
$
\Delta_t\odot M_t^{\mathrm{new}}
=
(\ell_t r_t^\top)\odot
\left(m_t^{\mathrm{new},L}(m_t^{\mathrm{new},Q})^\top\right)
=
(\ell_t\odot m_t^{\mathrm{new},L})
(r_t\odot m_t^{\mathrm{new},Q})^\top$.
Thus, New modulation can be viewed as redefining the effective write vectors, while the instantaneous update remains rank one. In contrast, Old modulation acts on \(\Theta_{t-1}\), which already contains information accumulated from previous inputs. The term \(\Theta_{t-1}\odot M_t^{\mathrm{old}}\) therefore directly controls the temporal memory state. This history-dependent memory control cannot be easily reproduced by only modifying the current update \(\Delta_t\), since \(\Delta_t\) is generated from the current input whereas \(\Theta_{t-1}\) summarizes the past.

\paragraph{Interpretation.}
The above derivation suggests that Old modulation captures the dominant mechanism of Self-Modulating QFWP: it determines how past quantum fast-weight updates are retained, suppressed, amplified, or sign-reversed through the temporal kernel \(K_{s\to t,j}^{\mathrm{old}}\). New modulation can still improve performance by refining the amplitude of newly written updates, which explains why the full model may perform best. However, when the task is primarily limited by memory dynamics rather than instantaneous write scaling, the Only-Old variant can naturally approach the performance of the full model.

\section{Conclusion}
In this work, we proposed Self-Modulating QFWP, a compact quantum fast-weight framework that adaptively regulates both newly generated updates and accumulated fast-weight memory. Across five time-series benchmarks and extensive hidden-size/sequence-length settings, the proposed model consistently improves convergence stability and predictive accuracy over standard QFWP. Our ablation studies further reveal that old-state modulation is the dominant source of improvement, indicating that effective control of historical quantum fast weights is central to robust sequential learning. These results position self-modulation as a simple but powerful mechanism for building adaptive and scalable quantum sequence models.

\clearpage
\bibliographystyle{IEEEtran}
\bibliography{bib/qml_examples,bib/fwp,bib/qfwp,bib/lstm,bib/qlstm,bib/qc,bib/qml_basics}

\end{document}